\documentclass[runningheads]{llncs}
\usepackage{graphicx}
\usepackage{amsmath}
\newtheorem{assumption}{Assumption}
\newtheorem{requirement}{Requirement}
\usepackage{bm} 
\usepackage{xcolor} 
\usepackage[colorlinks=true,linkcolor=blue]{hyperref}

\newcounter{myctr}

\definecolor{stringColor}{rgb}{0.8,0.1,0.1}
\usepackage{listings}
\usepackage{pxfonts}
\lstdefinelanguage{JavaScript}{
  keywords={typeof, new, true, false, catch, function, return, null, catch, switch, var, if, in, while, do, else, case, break, setTimeout, setInterval},
  keywordstyle=\color{black}\bfseries,
  ndkeywords={class, export, boolean, throw, implements, import, this},
  ndkeywordstyle=\color{darkgray}\bfseries,
  identifierstyle=\color{black},
  sensitive=false,
  comment=[l]{//},
  morecomment=[s]{/*}{*/},
  commentstyle=\color{stringColor}\ttfamily,
  stringstyle=\color{stringColor}\ttfamily,
  morestring=[b]',
  morestring=[b]"
}
\lstdefinelanguage{LF}{
  keywords={deadline, after, state, logical, physical, startup, shutdown, reaction, preamble, target, reactor, trigger, input, output, constructor, new, action, clock, actor, handler, time, main, federated, timer, sec, secs, msec, msecs, usec, usecs},
  emph={L,name, type, init, effect, instance}, emphstyle=\itshape,
  keywordstyle=\color{black}\bfseries,
  ndkeywords={class, export, boolean, throw, implements, import, this},
  ndkeywordstyle=\color{darkgray}\bfseries,
  identifierstyle=\color{black},
  sensitive=false,
  comment=[l]{//},
  morecomment=[s]{/*}{*/},
  commentstyle=\color{purple}\ttfamily,
  stringstyle=\color{black}\ttfamily,
  morestring=[b]',
  morestring=[b]"
}
\definecolor{borderColor}{rgb}{0.10,0.10,0.10}
\definecolor{historyColor}{rgb}{0.95,0.96,0.97}

\usepackage{xspace}
\newcommand{\lfshort}[0]{\textsc{LF}\xspace}

\newcommand{\lf}[0]{\textsc{Lingua Franca}\xspace}

\lstdefinestyle{framed} {
  xleftmargin=15pt,
  frame=l,
  basicstyle=\scriptsize\ttfamily,
  framesep=5mm,
  fillcolor=\color{gray!10},
  rulecolor=\color{gray!10},
  numberstyle=\normalfont\tiny\color{gray!90}
}
\usepackage{chngcntr}

\newcommand{\physicalConn}{\texttt{{\hbox{$\small\mathtt{\sim}$}}>}\xspace}

\begin{document}
\counterwithout{lstlisting}{section} 
\title{Quantifying and Generalizing the CAP Theorem}
%
%
\author{
Edward A. Lee\inst{1}\orcidID{0000-0002-5663-0584} \and
Soroush Bateni\inst{2}\orcidID{0000-0002-5448-3664} \and \\
Shaokai Lin\inst{1}\orcidID{0000-0001-6885-5572} \and
Marten Lohstroh\inst{1}\orcidID{0000-0001-8833-4117} \and
Christian Menard\inst{3}\orcidID{0000-0002-7134-8384}
}
\authorrunning{Lee et al.} 

\institute{UC Berkeley, Berkeley, CA, USA\\
\email{eal@berkeley.edu},
\email{shaokai@berkeley.edu},
\email{marten@berkeley.edu}
\and
UT Dallas, Richardson, TX, USA\\
\email{soroush@utdallas.edu}
\and
TU Dresden, Dresden, Germany\\
\email{christian.menard@tu-dresden.de}
}
\maketitle              
\begin{abstract}
In distributed applications, Brewer's CAP theorem tells us that when networks
become partitioned, there is a tradeoff between consistency and availability.
Consistency is agreement on the values of shared variables across a system, and
availability is the ability to respond to reads and writes accessing those shared
variables.  We quantify these concepts, giving numerical values to inconsistency
and unavailability. Recognizing that network partitioning is not an
all-or-nothing proposition, we replace the P in CAP with L, a numerical measure
of apparent latency, and derive the CAL theorem, an algebraic relation between
inconsistency, unavailability, and apparent latency.  This relation shows that
if latency becomes unbounded (e.g., the network becomes partitioned), then one
of inconsistency and unavailability must also become unbounded, and hence the
CAP theorem is a special case of the CAL theorem. We describe two distributed
coordination mechanisms, which we have implemented as an extension of the \lf 
coordination language, that support arbitrary tradeoffs between
consistency and availability as apparent latency varies.  With \emph{centralized
coordination}, inconsistency remains bounded by a chosen numerical value at the
cost that unavailability becomes unbounded under network partitioning.  With
\emph{decentralized coordination}, unavailability remains bounded by a chosen
numerical quantity at the cost that inconsistency becomes unbounded under
network partitioning. Our centralized coordination mechanism is an extension of
techniques that have historically been used for distributed simulation, an
application where consistency is paramount.  Our decentralized coordination
mechanism is an extension of techniques that have been used in distributed
databases when availability is paramount.
\end{abstract}
%
%
\section{Motivation} \label{sec:motivation}

In the year 2000, Eric Brewer gave a keynote talk~\cite{Brewer:00:CAP}
at the Symposium on Principles of Distributed Computing (PODC) in which he
introduced the ``CAP Theorem,'' which states that you can have only two of the following three properties in a distributed system:
\begin{itemize}
\item \textbf{Consistency}: Distributed components agree on the value of shared state.
\item \textbf{Availability}: Ability to respond to user requests.
\item tolerance to network \textbf{Partitions}: The ability to keep operating when communication fails.
\end{itemize}
This keynote is credited by many in the distributed computing community with opening up the field,
enabling innovative approaches that offer differing tradeoffs between these properties.

In 2002, Gilbert and Lynch \cite{GilbertLynch:02:CAP} proved a couple of variants of this theorem,
one rather strong result for ``asynchronous networks'' (which have no clocks) \cite[chapter 8]{Lynch:96:IOAutomata}
and one weaker result for ``partially synchronous networks'' (which have unsynchronized clocks that measure the passage of time
at the same rate).  These results reinforced Brewer's original interpretation that you could obtain any two of the three
properties, but not all three.

In 2012, Brewer wrote a retrospective~\cite{Brewer:12:CAP} in which he observes that tradeoff is more subtle than this. 
First, the ``P'' property is not really one you can trade off against the others.
You cannot choose whether to have network partitions.
Brewer clarified the design problem as one of how to trade off consistency against availability when network partitions occur.

In the same year, Abadi~\cite{Abadi:12:CAP} reacted that CAP is irrelevant when there are no network partitions,
but  Brewer~\cite{Brewer:12:CAP} points out that network partitions are not a binary property;
all networks have latency, and a complete communication failure is just the limiting case when the latency goes to infinity.
Abadi~\cite{Abadi:12:CAP}, in fact, emphasizes the tradeoff in database design between consistency and latency experienced by the user,
arguing that it has had more impact on actual database system designs than CAP.
He then observes that ``availability and latency are arguably the same thing'' 
and calls for unifying the consistency/latency tradeoff with the CAP tradeoffs.

There are (at least) two latencies of interest: network latency and user latency.
Whereas \textbf{network latency} is the time that it takes messages to traverse the network, user latency, which 
we will refer to as \textbf{unavailability}, is
the time it takes to respond to a user.
Specifically, unavailability is the physical time that elapses
between when a system first receives a request from a user and when the system sends a response to the user.
Network partitioning is the limiting case of network latency.
When the network latency goes to infinity, the network is partitioned.
Hence, we derive a \textbf{CAL} theorem,
replacing ``tolerance for network Partitioning'' with ``tolerance for network Latency.''
This theorem will establish a numerical relationship between inconsistency, unavailability, and network latency,
thereby unifying Brewer's and Abadi's positions and having the CAP theorem as a special case.
Our resulting formulas make clear the tradeoff between these quantities
and enable a range of designs that vary in their emphasis on one or more of these quantities.

To establish the relationship, we have to carefully define the three variables:
inconsistency, unavailability, and network latency.
Most designers have strong intuitive understandings of the meanings of these terms,
but there are pitfalls.
For example, an intuitive statement about a consistent view of a shared variable $x$ goes something like this:
If component 1 writes to $x$ ``before'' component 2
reads the value of $x$, then component 2 will read the updated value.
Another such statement is that if component 1 writes to $x$ ``before'' component 2 also writes to $x$, then all observers of $x$
will ``subsequently'' see the value written by component 2.
These statements are difficult to formalize, however, because the words ``before'' and ``subsequently'' are problematic.
The order in which two physically separated events occur may be not only unknown,
but also, at a fundamental level, \textbf{unknowable}.
From special relativity, the order in which physically separated events occur depends on
the frame of reference.
In modern distributed systems---some of which span the globe and all of which have communication
that is limited by the speed of light---if many updates and reads of shared variables are occurring,
some of those are sure to have ambiguous ordering.
So we will have to do better than this.
We need a more solid foundation for the notion of consistency.

In Section~\ref{sec:consistency}, we give an overview of the concept of consistency, the role of time, and the use of timestamps.
In Section~\ref{sec:cal}, we define the numerical quantities inconsistency, unavailability, and network latency
and derive the CAL theorem, which relates these three quantities.
In Section~\ref{sec:example}, we give complete programs illustrating the tradeoffs implied by the CAL theorem,
and show how the \lf coordination language supports being explicit about these tradeoffs.
In Section~\ref{sec:implementation}, we describe two implementations: one that is capable of bounding inconsistency,
and one that is capable of bounding unavailability in the face of network partitions.
Section~\ref{sec:conclusion} offers some concluding remarks.

\section{Consistency and Time}\label{sec:consistency}

Consider a shared variable $x$ that is read and written to by various sequential processes running on nodes distributed across a network.
An intuitive form of consistency requires that all nodes agree on the order of operations on $x$ (reads and writes).
Gilbert and Lynch say ``The most natural way of formalizing the idea of a consistent service is as an atomic data object''
\cite{GilbertLynch:02:CAP},
where ``atomic'' comes from Lamport \cite{Lamport:86:Distributed}.
``Atomic'' here is
interchangeable with the term ``linearizable'' that comes from Herlihy and Wing \cite{HerlihyWing:90:Linearizable}.
This interpretation defines ``consistent'' to mean that all observers see the same sequence of atomic operations
(reads and writes) on the data object.

For most applications, however, this is too strong.
It is sufficient to agree on the order in which writes occur, and to agree on the order in which reads occur relative to those writes,
but otherwise, there is usually no need to agree on the order in which the reads occur relative to each other.
But even this weaker agreement may be too strong.
Consider two independent writes to a shared variable $x$ that occur nearly simultaneously on two distinct nodes.
Depending on how these writes are to be merged, it may not be necessary to define the order in which they occur.

There is also the question of what we mean by ``agreement.''
When should the nodes agree?
And what is agreement?

These questions have as a backdrop some notion of physical time.
Intuitively, physical time has to play a major role.
It would be strange to design a system that considers a write to $x$
that occurs three weeks after another write to $x$ to have occurred ``before.''
But physical time is a slippery, poorly understood thing.
Newtonian time, with all its beauty and simplicity, has limited applicability, and there is no perfect measurement of it.
Fundamentally, the ordering and agreement that we need are logical (or semantic) properties, not physical properties.
But to make them useful in applications, we will need physical realizations that are reasonably faithful to our
semantic models using only practical, realizable measurements of physical time.
We require, therefore, formal semantic models
and physical realizations that, with high confidence, conform to these models.

The events we will consider in this paper are reading and writing values to or from some shared variable $x$.
The value of $x$ may be a text, a number, or some data structure,
and the semantics of a ``write'' will matter.
In general, a ``write'' is a \textbf{merge} of a new value with an old value.
For example, $x$ may be a number and a ``write'' may \textbf{replace} the old value
or simply \textbf{add} to it.
The latter merge operation is associative and commutative, assuming perfect arithmetic, where there are no overflows or rounding errors.
The replacement merge operation is associative but not commutative.
If $x$ is a text, the ``write'' may \textbf{append} to the text, again giving a merge operation that is also associative but not commutative.

If the merge operation is associative and commutative, then we have two of the three properties of
the ACID 2.0 database principle, proposed by Helland and Campbell \cite{HellandCampbell:09:ACID2}. 
The letters stand for Associative, Commutative, Idempotent, and Distributed.
Idempotence means that operations that are applied more than once have the same effect as operations that are applied exactly once.
We will assume that idempotence is realized at a layer below, not in the application code.
In Section~\ref{sec:example}, we describe a system realization where this is the case.
Hence, the application code can assume that writes are applied exactly once (in the absence of failures).

If we are to have availability in the presence of network partitioning, then some amount of \textbf{replication} is necessary~\cite{Abadi:12:CAP}.
That is, the value of $x$ will need to be stored at multiple nodes.
It is this replication that creates the consistency problem, but without it, we cannot assure availability. 
We will assume that reads and writes to $x$ can occur at any of the nodes.

\subsection{Causal Consistency}

An elegant formal model of consistency, called \textbf{causal consistency},
is analyzed in depth by Schwartz and Mattern \cite{SchwarzMattern:94:CausalConsistency}.
They define a causality relation, written $e_1 \rightarrow e_2$, between events $e_1$ and $e_2$ to mean that $e_1$ can causally affect $e_2$.
The phrase ``causally affect'' is rather difficult to pin down (see Lee \cite[Chapter 11]{Lee:20:Coevolution} for the subtleties around the notion of causation),
but, intuitively, $e_1 \rightarrow e_2$ means $e_2$ cannot behave as if $e_1$ had not occurred.
Put another way, if the effect of an event is reflected in the state of a local replica of a variable $x$, then any cause of the event
must also be reflected. Put yet another way, an
observer must never observe an effect before its cause.

For example, if $e_1$ writes $x = 1$ and $e_2$ sends the value of $x$ to another process, then $e_1 \rightarrow e_2$ means that whatever value $e_2$ sends is either the $x = 1$ value written by $e_1$ or some subsequently written value.

Following Schwartz and Mattern, we will impose a bit more structure on our distributed applications.
First, we define a \textbf{process} to be a totally ordered sequence of events occurring on a single machine 
and maintaining a single replica of one or more shared variables.
A process, therefore, is the trace (the record of events) of a single thread of execution.
Second, for each shared variable $x$, we will consider four types of events:
\begin{enumerate}
\item $w_x$: Merge a value with the local replica of $x$.
\item $d_x$: Read the value of the local replica of $x$.
\item $s_x$: Send the value of the local replica of $x$ to some set of other processes.
\item $r_x$: Receive a new value for $x$ and merge it with the local replica.
\end{enumerate}
When the type of event is important, we will use one of the four symbols above to denote the event.
Otherwise, we will use the symbol $e$.
If the variable $x$ is clear from context, we will omit the subscript.
Typical merge operations are \textbf{replace}, \textbf{append}, and \textbf{add}, but the merge operation can be arbitrarily sophisticated.
To maintain consistency, we will find it necessary for all replicas of $x$ to use the same merge operation.

Formally, the causality relation of Schwartz and Mattern is the smallest \emph{transitive} relation such that
$e_1 \rightarrow e_2$ if $e_1$ precedes $e_2$ in a process,
or $e_1$ is the sending of a value in one process (event type $s_x$) and $e_2$ is the receipt of the value in another process (event type $r_x$).
If neither $e_1 \rightarrow e_2$ nor $e_2 \rightarrow e_1$ holds, then we write $e_1 || e_2$ or $e_2 || e_1$ and say that $e_1$ and $e_2$ are \textbf{incomparable}.
The causality relation is identical to the ``happened before'' relation of Lamport \cite{Lamport:78:Time},
but Schwartz and Mattern prefer the term ``causality relation''
because even if $e_1$ occurs unambiguously earlier than $e_2$ in physical time,
they may nevertheless be incomparable, $e_1 || e_2$.

The causality relation is a strict partial order.
Schwartz and Mattern use their causality relation to define a ``consistent global snapshot'' of a distributed computation
to be a subset $S$ of all the events $E$ in the execution that is a downset,
meaning that if $e' \in S$ and $e \rightarrow e'$, then $e \in S$
(this was previously called a ``consistent cut'' by Mattern~\cite{Mattern:88:virtualtime}).

\begin{figure}[tb]
\centering
\includegraphics[width=3in]{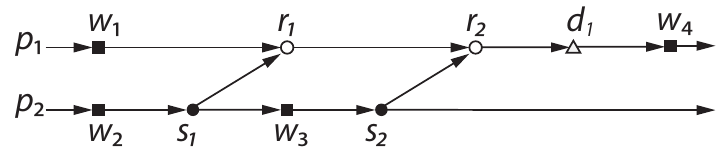}
\caption{Causal ordering of events in a bulletin board example. \label{fig:causal}}
\end{figure}

Consider for example a shared online bulletin board, like a Facebook Timeline.
Consider the following events, adapted from Bailis et al. \cite{BailisEtAl:13:CausalConsistency}:
\begin{itemize}
\item $w_1$: Joe posts a picture of Sally at a recent party by writing to a local copy.
\item $w_2$: Sally posts that her son Billy is missing, writing to a local copy.
\item $s_1$: Sally's message is sent to Joe's process.
\item $r_1$: Joe's machine receives the message and updates his local copy.
\item $w_3$: Sally posts that her son has been found (on the local copy).
\item $s_2$: Sally's message is sent to Joe's process.
\item $r_2$: Joe's machine receives the second message and updates his local copy.
\item $d_1$: Joe reads Sally's messages.
\item $w_4$: Joe posts ``That's good news, a relief.''
\end{itemize}
The immediate causal relations are depicted by arrows in Fig.~\ref{fig:causal}.
The complete causal relation is the transitive closure of the depicted causal relations.

\begin{figure}[tb]
\centering
\includegraphics[width=4in]{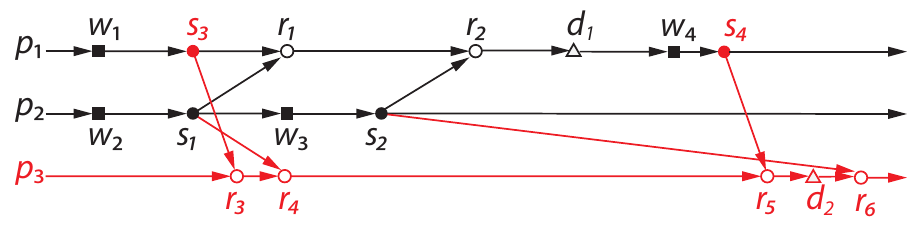}
\caption{Observer of the exchange between Joe and Sally. \label{fig:observer}}
\end{figure}

Suppose that in this application the merge operation is \textbf{append}.
Each write to the local copy of the bulletin board simply appends the message to the end of the board.
This operation is associative but not commutative.
After the writes $w_1$ and $w_2$, Sally and Joe have inconsistent views of the shared variable.
A third observer, as indicated in Fig.~\ref{fig:observer}, may have yet a third inconsistent view.
In the figure, after the event  $r_4$, the $p_3$ observer sees both messages, albeit in arbitrary order that has been determined by the happenstances of network latency.
However, in this case, there is no violation of \emph{causal} consistency because $w_1 || w_2$.
Semantically, Joe's and Sally's first writes are unordered, and different observers may see none, one, or both messages, and in the latter case, they may see them in either order.
Causal consistency tolerates this form of nondeterminism.

A worse problem, however, is illustrated later in the same figure.
After event $r_5$, the $p_3$ observer has seen Sally's first post and Joe's second (``That's good news'')
but has not seen Sally's second post.
At the read event $d_2$, this observer will think that Joe has responded ``that's good news''
to Sally's message that Billy is missing!
This observer would have to conclude that Joe really does not like Sally's son.
After event $r_5$, the state of $p_3$'s local replica is \emph{causally inconsistent} because $p_3$ sees the effects of $w_4$,
but not the effects of $w_3$, and $w_3 \rightarrow w_4$.

Leveraging the results of Charron-Bost~\cite{Charron-Bost:91:VectorClocks}, Schwartz and Mattern argue
that there is no simple solution that ensures causal consistency that does not also impose additional unnecessary constraints.
They give a non-simple solution using \textbf{vector clocks}.
It is non-simple in that it requires, in the worst case, message sizes that are order $N$, where $N$ is the number of communicating processes.
In subsequent sections, we will describe a solution using timestamps that has constant overhead and exploits static information about 
communication patterns between nodes
in order 
to reduce the number of unnecessary additional constraints.

\subsection{Causal Consistency is Insufficient}

In practice, causal consistency is necessary but not sufficient for most applications.
It may be perfectly reasonable to impose additional constraints.
In the bulletin board application, for example, it may not be acceptable for different observers to see Sally and Joe's first posts in arbitrary order.
One common way to prevent that is to timestamp the posts using some clock measuring physical time on the machine where the post originates.
All observers could then arrange whatever posts they see in timestamp order, breaking ties with some consistent algorithm (e.g. alphabetical ordering).
These timestamps give a semantics to the problematic ``before'' and ``subsequently'' that we encountered in Section~\ref{sec:motivation}.
A general form of this solution is often called ``last writer wins,'' but that phrase only makes sense if the merge policy is \textbf{replace}.
So we will call the policy \textbf{timestamp-ordered}.

Some care is required here because unsynchronized clocks could result in rather unexpected orderings.
Hence, this solution assumes some measure of clock synchronization, a topic we discuss further in Section~\ref{sec:clocks}.
Nevertheless, this solution is fairly common, found in many distributed database systems, such as Google Spanner~\cite{CorbettEtAl:12:Spanner,CorbettEtAl:13:Spanner}.

By itself, however, time-stamping events is not enough to ensure causal consistency.
After $r_5$ in Fig.~\ref{fig:observer}, $w_3$ will still be missing even after the posts seen are arranged in timestamp order.
Ensuring causal consistency will require more work.
We describe two mechanisms in Section~\ref{sec:implementation} that accomplish this.

An additional requirement in many applications is \textbf{eventual consistency}.
Eventual consistency, informally, means that if writes stop occurring at some point in time, then, eventually, all replicas of $x$ will have the same value.
Fig.~\ref{fig:observer} would become eventually consistent if we add send-receive pairs so that $p_1$ sends its writes to $p_2$.
However, even though it will be eventually consistent, it will still not be causally consistent unless something is done about the mis-ordered $r_5$ and $r_6$.

We will henceforth assume that our distributed system is \textbf{causally} and \textbf{eventually consistent} with \textbf{timestamps},
which we shorten to \textbf{CET}.
The first and last properties are absolute in that they hold at all times during an execution.
Eventual consistency, however, is rather odd in that an eventually consistent system may, in fact, never be consistent!
It may become consistent only if all processes stop writing to shared variables,
and even then, it may take an unbounded amount of time to become consistent.

We will formally define a time interval $\bar{C}$ (technically, a timestamp interval) over which a system is allowed to be inconsistent.
$\bar{C}$ will become a measure of \textbf{inconsistency} in the system, and we will give relations between $\bar{C}$, a measure $\bar{A}$ of \textbf{unavailability},
and a measure $\mathcal{L}$ of (apparent) \textbf{network latency}.
If $\bar{C}$ is bounded, then the system is eventually consistent.
If $\bar{C}$ is zero, we call the system \textbf{strongly consistent}.
A strongly consistent system is obviously eventually consistent.

\subsection{Using Physical Time to Assign Logical Times} \label{sec:clocks}

To replace the naive ``earlier'' with something realizable,
we borrow an idea from Lamport~\cite{Lamport:84:TimeStamps},
who proposed that each update of a shared variable be assigned a timestamp.
The timestamp is based on a local clock, and the local clocks throughout the system are sufficiently synchronized to provide
an application-dependent reasonable notion of ``before'' or ``earlier.''
Once a timestamp is assigned, it is treated as a \textbf{logical time} and defines the semantic order of events.

In the bulletin board example, the timestamps
could be based on Coordinated Universal Time (UTC) and synchronized via the Network Time Protocol (NTP)  \cite{Mills:03:NTP,Mills:06:NTP}, which is designed to
synchronize computer clocks to within a few milliseconds of UTC.
NTP is ubiquitous in modern computers, so this is not likely to present a practical problem.
If a post in London gets a timestamp 10 milliseconds less than a post in Singapore, the London post will be deemed to have occurred before.
Whether it physically occurred before the post in Singapore is irrelevant.\footnote{In fact, the physical order of these two events is unknowable. At the speed of light, Singapore is about 32 milliseconds away from London, which means that even with perfect clock synchronization, these two events are outside each other's light cones. This implies that the physical order of the two events will be different in different frames of reference.}

Given such timestamped events,
the behavior of the distributed system can then be defined by the numerical order of those timestamps.
Once the timestamps are assigned, their numerical order is a semantic property, and the goal of a CET system design
is to ensure that every component sees events in timestamp order.
This is the semantics adopted by Google Spanner~\cite{CorbettEtAl:12:Spanner,CorbettEtAl:13:Spanner},
a globally distributed replicated database operating today.

Because the timestamps are numbers, there is always the possibility of two events having the same timestamp.
These will be deemed to have occurred ``simultaneously.''
How to handle simultaneous events is application dependent.
If the merge operation is associative and commutative, then the simultaneous operations can be applied in different orders without adverse consequences.
If, further, no user can observe the intermediate results of these operations, then the whole system will be strongly consistent.
Without associative and commutative operations, however, the order of operations will need to be the same at all nodes in order to achieve
consistency.  For simultaneous events, consistent ordering
requires a consistent mechanism for assigning distinct priorities to each event.

Note that even sloppy clock synchronization can be useful because it defines a semantic ordering.
In the bulletin board application, does it really matter if the clocks are off even by a few seconds (which would be unusually poor performance for NTP)?
We will see that clock synchronization error is indistinguishable from network latency, and so the sum of the two quantities is
the only meaningful measure.
Consequently, a large clock synchronization error coupled with a requirement for strong consistency will result in high unavailability.
This would be true with any mechanism for assigning a total order to distributed updates because any such order,
in effect, assigns some kind of logical ``time'' to each event.
The mere act of putting them in order is isomorphic with assigning them a time.

On local area networks and in data centers, much higher precision clock synchronization is available with PTP (precision time protocol)
\cite{Eidson:06:1588,EidsonStanton:15:Time},
a centerpiece of the TSN (time sensitive network) family of protocols being developed by the IEEE 802.1 working group \cite{LoBelloEtAl:19:TSN}.
The clock synchronization standard part of TSN, IEEE 802.1AS-2020, a subset of the older standard IEEE 1588, is widely supported in networking hardware,
even if it has not yet appeared in standard programming APIs.
This standard is capable of synchronizing clocks to within less than a nanosecond on local area networks, and on the order of microseconds on wide area networks.
GPS (the global positioning system) provides a complementary alternative that is sometimes combined with PTP.
With GPS, clocks can be synchronized to sub-nanosecond precision globally, in principle \cite{JeffersonEtAl:96:GPS}.
Google Spanner synchronizes master clocks at data centers using GPS combined with atomic clocks,
and then synchronizes clocks within the data center using PTP~\cite{CorbettEtAl:12:Spanner}.

\section{The CAL Theorem} \label{sec:cal}

We can now (finally!) define the terms in the CAL theorem,
\textbf{inconsistency}, \textbf{unavailability}, and (apparent) \textbf{latency}.
As we will see, network latency is only one of several factors that can affect the other two quantities, so
we drop the reference to the ``network'' and simply consider ``latency,''
trusting that the reader will not apply the word ``latency'' to ``unavailability,'' as Abadi does~\cite{Abadi:12:CAP}.
We will see that network latency is indistinguishable from clock synchronization error,
and that computation times can have just as detrimental effects as network latency.
Hence, we lump all these three effects, network latency, clock synchronization error, and execution times under one heading,
``latency,'' or ``apparent latency'' when there might be some ambiguity about which latency we are talking about.

\subsection{Tags and Physical Time}\label{sec:tags}

We are interested in times of events and time intervals between events, but
we have to be careful because these cannot simply be Newtonian time.
Newtonian time is inaccessible to distributed software.
We will instead use two distinct notions of time, logical and physical.
A physical time $T \in \mathbb{T}$ will be an imperfect measurement of time 
taken from some clock somewhere the system.

The set $\mathbb{T}$ contains all the possible times that a physical clock can report.
We assume that $\mathbb{T}$ is totally ordered and includes two special members:
$\infty \in \mathbb{T}$ is larger than any time any clock can report, and
$-\infty \in \mathbb{T}$ is smaller than any time any clock can report.
For example, $\mathbb{T}$ could be the set of integers $\mathbb{Z}$ augmented with the two infinite members.

Given any $T_1, T_2 \in \mathbb{T}$, the \textbf{physical time interval} 
(or just \textbf{time interval} if there is no ambiguity) between the two times is written $i = T_1 - T_2$.
Time intervals are assumed to be members of a group $\mathbb{I}$ with a largest member
$\infty$ and smallest member $-\infty$ and a commutative and associative addition operation.
For example, $\mathbb{I}$ could be the set of integers $\mathbb{Z}$ augmented with the two infinite members.
Addition involving the infinite members behaves
in the expected way in that for any $i \in \mathbb{I} \setminus \{\infty, -\infty\}$,
\begin{eqnarray*}
i + \infty &=& \infty \\
i + (- \infty) &= &-\infty \\
\end{eqnarray*}
We also assume that addition of infinite intervals saturates, as in
\begin{eqnarray*}
\infty + \infty &=& \infty \\
(-\infty) + (- \infty) &= &-\infty \\
\infty + (- \infty) && \text{is undefined}
\end{eqnarray*}
Note that we use the same symbols $\infty$ and $-\infty$ for the special members of both the set of physical times $\mathbb{T}$
and the set of intervals $\mathbb{I}$. We hope this will not create confusion.

Intervals can be added to a physical time value, and we assume that this addition is associative.
I.e., for any $T \in \mathbb{T}$ and any $i_1, i_2 \in \mathbb{I}$,
\begin{equation}
T + (i_1 + i_2) = (T + i_1) + i_2 \in \mathbb{T} .
\end{equation}
Addition of infinite intervals to a time value saturates in a manner similar to addition of infinite intervals.

These idealized requirements for physical times and time intervals can be efficiently approximated in practical implementations.
First, it is convenient to have the set $\mathbb{T}$ represent a common definition of physical time, such as Coordinated Universal Time (UTC)
because, otherwise, comparisons between times will not correlate with physical reality.
In the \lf language that we use in Section~\ref{sec:example}, $\mathbb{T}$ and $\mathbb{I}$ are both
the set of 64-bit integers.
A $T \in \mathbb{T}$ is a POSIX-compliant representations of time,
where $T$ represents the number of nanoseconds that have elapsed since midnight, January 1, 1970, Greenwich mean time.
In the \lf realization, the largest and smallest 64-bit integers represent $\infty$ and $-\infty$, respectively,
and addition and subtraction respect the above saturation requirements.
Note, however, the set of 64-bit integers is not the same as the set $\mathbb{Z}$ because it is finite.
As a consequence, addition can overflow.
In \lf, such overflow saturates at $\infty$ of $-\infty$, and as a consequence, addition is no longer associative.
For example, $T + (i_1 + i_2)$ may not overflow while $(T + i_1) + i_2$ does overflow.
As a practical matter, however, this will only become a problem with systems that are running near the year 2270.
Only then will the behavior deviate from the ideal given by our theory.

For \emph{logical} time, we use an element that we call a \textbf{tag} $g$ of a totally-ordered set $\mathbb{G}$.
Each event in a distributed system is associated with a tag $g \in \mathbb{G}$.
From the perspective of any component of a distributed system, the order in which events occur is defined by the order of their tags.
If two distinct events have the same tag, we say that they are \textbf{logically simultaneous}.
We assume the tag set $\mathbb{G}$ has an element $\infty$ that is larger than any other tag and another $-\infty$ that is smaller than any other tag.

In the \lf language,
$\mathbb{G} = \mathbb{T} \times \mathbb{U}$, where $\mathbb{U}$ is the set of 32-bit unsigned integers representing the microstep of a superdense time system~\cite{Maler:92:Hybrid,Cataldo:06:Tetric,CremonaEtAl:17:Hybrid}.
We use the term \textbf{tag} rather than timestamp to allow for such a richer model of logical time.
For the purposes of this paper, however, the microsteps will not matter, and hence you can think of a tag as a timestamp and ignore the microstep.
We will consistently denote tags with a lower case $g \in \mathbb{G}$ and measurements of physical time $T \in \mathbb{T}$ with upper case.

We will need operations that combine tags and physical times.
To do this, we assume a monotonically nondecreasing function $\mathcal{T}\colon \mathbb{G} \to \mathbb{T}$
that gives a physical time interpretation to any tag.
For any tag $g$, we call $\mathcal{T}(g)$ its \textbf{timestamp}.
In \lf, for any tag $g = (t,m) \in \mathbb{G}$, $\mathcal{T}(g) = t$.
Hence, to get a timestamp from a tag, you just have to ignore the microstep.

The set $\mathbb{G}$ also includes infinite elements such that
$\mathcal{T}(\infty_{\mathbb{G}}) = \infty_{\mathbb{T}}$
and
$\mathcal{T}(-\infty_{\mathbb{G}}) = -\infty_{\mathbb{T}}$,
where the subscripts disambiguate which infinity we are referring to.

An external input from outside the system, such as a user input or query, will be assigned a tag $g$ such that
$\mathcal{T}(g) = T$,
where $T$ is a measurement of physical time taken from the local clock where the input first enters the system.
In \lf, this tag is normally given microstep 0, $g = (T,0)$.


\subsection{Inconsistency}\label{sec:inconsistency}

We will define \textbf{inconsistency}, $\bar{C}$, to be a non-negative time interval in $\mathbb{I}$
that has the value zero for strongly consistent systems.
When it has a value greater than zero, it will represent a time interval over which components in a distributed system
can disagree about the values assigned to a shared data object.
If this number is bounded, then we have \textbf{eventual consistency}.

Assume we are given a \textbf{trace} of an execution of a distributed system.
As before, we assume the distributed system consists of $N$ sequential processes,
and each sequential process is a potentially unbounded sequence of (tagged) \textbf{events}.
The $k$-th event
is associated with a tag $g_k$ and a physical time $T_k$.
The physical time $T_k$ is the reading on a local clock at the time where the event starts being processed.

The events in a process are required to have nondecreasing tags and increasing physical times.
That is, if $g_k$ is the tag and $T_k$ is the physical time of the $k$-th event, then
$g_k \le g_{k+1}$ and $T_k < T_{k+1}$.
The constraint that a process have nondecreasing tags is central to maintaining causal consistency,
but it comes with significant costs.
We will show in Section~\ref{sec:implementation} how a practical runtime infrastructure can enforce this requirement.

The constraint that physical times be increasing is, perhaps surprisingly to some readers, not trivial to ensure.
The physical time of an event is the reading of a local clock when the processing of the event starts.
But local clocks are not always assured of increasing monotonically.
A clock that is synchronized using NTP, for example, can move backward when updated.
The \lf realization used in Section~\ref{sec:implementation} provides an interface to the system clock that is guaranteed to be increasing.

A \textbf{write event} is an update to a local copy of a shared variable.
A \textbf{read event} is a reading of the local copy of a shared variable.
A \textbf{send event} is the launching into the network of a notification of a write event;
we assume that, in the trace, every write event is followed by a corresponding send event.
A \textbf{receive event} is the receipt of such a notification;
we assume that every receive event is also a write event that updates the local copy of the shared
variable with the value received.
All updates are mediated by some application-specific \textbf{merge operation} that
combines the update value with the previous value in some specified way.

Within each process,
every read event with tag $g$ yields the value of a shared variable $x$ that was assigned to the local copy of $x$
by the write or receive event in the same process with the largest tag $g'$ where $g' \le g$.
If $g' = g$, we require that $T' < T$, where $T'$ is the physical time of the write or receive event
and $T$ is the physical time of the read event.
This requirement ensures that a read event reads a value that was written at an earlier physical time.
This requirement is the principal reason for the requirement that physical times in a process be monotonically increasing.

We have additional requirements on tags:
\begin{requirement} \label{req:tags}
A send event has a tag greater than or equal to that of the write event that it is reporting
and a physical time greater than that of the write event.
A receive event has a tag greater than or equal to the tag of the send event that it is receiving.
The physical time of the receive event relative to the send event is unconstrained.
\end{requirement}

\begin{definition}\label{def:consistency}
For each write event on process $j$ with tag $g_j$,
let $g_i$ be the tag of the corresponding receive event on process $i$
or $\infty$ if there is no corresponding receive event.
The \textbf{inconsistency} $\bar{C}_{ij} \in \mathbb{I}$ from $j$ to $i$ is defined to be
\begin{equation}
\bar{C}_{ij} = \max(\mathcal{T}(g_i) - \mathcal{T}(g_j)),
\end{equation}
where the maximization is over all write events on process $j$.
If there are no write events on process $j$, then we define $\bar{C}_{ij} = 0$.
\end{definition}

With  requirement \ref{req:tags}, it is clear that $\bar{C}_{ij} \ge 0$.
If $\bar{C} = 0$,
we have \textbf{strong consistency}.
We will see that this strong consistency comes at a price in availability,
and that network failures can result in unbounded unavailability.
If $\bar{C}$ is bounded, we have \textbf{eventual consistency}, and $\bar{C}$ quantifies ``eventual.''

\subsection{Unavailability}

Unavailability, $\bar{A}$, is a measure of the time it takes for a system to respond to user requests~\cite{kleppmann2015critique}.
If it takes a long time (or it never responds), then the system is unavailable, whereas if responses
are instantaneous, then the system is highly available.

A user request is an external event that originates from outside the distributed system.
Assume that a user request triggers a read event in process $i$ with tag $g_i$ such that its timestamp $\mathcal{T}(g_i)$ is
the reading of a local clock when the external event occurs.
Let $T_i$ be the physical time of the read event, i.e., the physical time at which the read is processed.
Hence, $T_i \ge \mathcal{T}(g_i)$.
\begin{definition}\label{def:availability}
For each read event on process $i$, let $g_i$ be its tag and $T_i$ be the physical time
at which it is processed. The \textbf{unavailability} $\bar{A}_i \in \mathbb{I}$ at process $i$ is defined to be
\begin{equation}
\bar{A}_i = \max(T_i - \mathcal{T}(g_i)),
\end{equation}
where the maximization is over all read events on process $i$ that are triggered by user requests.
If there are no such read events on process $i$, then $\bar{A}_i = 0$.
\end{definition}
Because we are considering only read events that are triggered by external user requests,
$\mathcal{T}(g_i) \le T_i$, so $\bar{A}_i \ge 0$.
If $\bar{A}_i = 0$, then we have maximum availability (minimum unavailability).
This situation arises when reads are immediately satisfied.
To maintain strong consistency across a distributed system, however, we will find that
unavailability is typically larger, $\bar{A}_i > 0$. 

\subsection{Processing Offsets}

To maintain causal consistency, we require that a process have nondecreasing tags.
For this reason, in a trace, a read or write event triggered by an external input
may have a physical time $T$ that is significantly larger than its tag's timestamp $\mathcal{T}(g)$.
While $\mathcal{T}(g)$ is determined by the physical clock at the time the external input appears,
the physical time at which the event is actually processed may have to be later to ensure that all events with earlier tags have been processed.

Here, it becomes convenient to assume that write events \emph{with the same tag} have an associative and commutative merge operation.
Consistency of any kind would not be possible were this not the case because identical tags cannot be used to determine in which order to apply updates.

A consequence of this assumption is that a process can process a \emph{write} event whenever it is sure that no events with \emph{earlier} tags will
later appear.
In contrast, a process can only process a \emph{read} event when it is sure that no events with \emph{earlier or equal} tags will later appear.
The result of the read should reflect \emph{all} updates to the shared variable with tags equal to or less than that of the read event.

This motivates the following definition:
\begin{definition}\label{def:processing}
For process $i$, the \textbf{processing offset} $O_i \in \mathbb{I}$ is
\begin{equation}
O_i = \max(T_i - \mathcal{T}(g_i))
\end{equation}
where $T_i$ and $g_i$ are the physical time and tag, respectively,
of a write event on process $i$ that is triggered by a local external input
(and hence assigned a timestamp drawn from the local clock).
The maximization is over all such write events in process $i$.
If there are no such write events, then $O_i = 0$.
\end{definition}
Notice that the processing offset closely resembles the unavailability of Definition~\ref{def:availability},
but the former refers to \emph{write} events and the latter refers to \emph{read} events.

The processing offset, by definition, is greater than or equal to zero.
We will see that it is often zero, but not always.

\subsection{Apparent Latency}

When a write to a shared variable occurs in process $j$, some time will elapse before a corresponding receive event
on process $i$ performs a corresponding write to its local copy of the shared variable.
This motivates the following definition:

\begin{definition}\label{def:apparent}
Let $g_j$ be the tag of a write event in process $j$ that is triggered by an external input at $j$
(so $\mathcal{T}(g_j)$ is the physical time of that external input).
Let $T_i$ be the physical time of the corresponding receive event in process $i$ (or $\infty$ if there is no such event).
(If $i=j$, we assume $T_i$ is the same as the physical time of the write event.)
The \textbf{apparent latency} or just \textbf{latency} $\mathcal{L}_{ij} \in \mathbb{I}$ for communication from $j$ to $i$ is
\begin{equation}\label{eq:apparent}
\mathcal{L}_{ij} = \max(T_i - \mathcal{T}(g_j)),
\end{equation}
where maximization is over all such write events in process $j$.
If there are no such write events, then $\mathcal{L}_{ij} = 0$.
\end{definition}

Note that $T_i$ and $\mathcal{T}(g_j)$ are physical times \emph{on two different clocks} if $i \neq j$,
so this apparent latency is an actual latency only if those clocks are perfectly synchronized.
Unless the two processes are actually using the same physical clock, they will never be perfectly synchronized.
Hence, the apparent latency may even be negative.
Note that despite these numbers coming from different clocks, if tags are sent along with messages, this apparent latency is measurable at the receiving end.

The apparent latency can be thought of as the sum of four components,
\begin{equation}\label{eq:latency}
\mathcal{L}_{ij} = O_j + X_{ij} + L_{ij} + E_{ij} ,
\end{equation}
where $X_{ij}$ is \textbf{execution time} overhead at node $j$ for sending a message to node $i$,
$L_{ij}$ is the \textbf{network latency} from $j$ to $i$, and $E_{ij}$ is the \textbf{clock synchronization error}.
The three latter quantities are indistinguishable and always appear summed together,
so there is no point in breaking apparent latency down in this way.
Moreover, these latter three quantities would have to be measured with some physical clock,
and it is not clear what clock to use.
The apparent latency requires no problematic measurement since it explicitly refers to local clocks and tags.

The clock synchronization error can be positive or negative, whereas $O_j$, $X_{ij}$, and $L_{ij}$ are always nonnegative.
If $E_{ij}$ is a sufficiently large negative number, the apparent latency will itself also be negative.
Because of the use of local clocks, the receive event will appear to have occurred before the user input that triggered it.
This possibility is unavoidable with imperfect clocks.

\subsection{The CAL Theorem for Strong Consistency}

We can now put everything together and identify the sources of the unavailability of Definition~\ref{def:availability}.
In that definition, $T_i$ must be greater than $\mathcal{T}(g_i)$ by enough to ensure that a read event at process $i$ is processed
after all writes to the shared variable with tags less than or equal to that of the read event.
From this, we can see that
\begin{equation}\label{eq:unsummed}
\bar{A}_i = \max_{j \in N}(\bar{A}_{ij}),
\end{equation}
where $\bar{A}_{ij}$ is the unavailability at $i$ if only the two processes $i$ and $j$ are considered.

Suppose that process $j$ has a write event with tag $g_j$ triggered by a local external event.
In the worst case, the receive event on $i$ corresponding to this write event will have physical time
\begin{equation}\label{eq:wait}
T_i = \mathcal{T}(g_j) + \mathcal{L}_{ij} ,
\end{equation}
from (\ref{eq:apparent}).
Note that $T_i$ is measured by the clock at $i$,
and, since $\mathcal{L}_{ij}$ can be negative (provided $i \neq j$), it is possible for $T_i$ to be less than $\mathcal{T}(g_j)$.

Node $i$ cannot respond to the local read query with tag $g_i$ before physical time reaches $\mathcal{T}(g_i) + O_i$
because $O_i$ is the time it takes to make sure all \emph{earlier} events have been processed.
But the possibility of incoming network messages means that \emph{additional} delay may be required
to ensure that earlier \emph{or equal} tags have been processed.
That is, if there is the possibility of a remote write event on $j$ with tag $\mathcal{T}(g_j) = \mathcal{T}(g_i)$, then
(\ref{eq:wait}) implies that $i$ needs to also wait at least until physical time
\begin{equation}\label{eq:wait2}
T'_i = \mathcal{T}(g_i) + \mathcal{L}_{ij} ,
\end{equation}
where $\mathcal{L}_{ij}$ includes $O_j$.
Therefore, to maintain strong consistency, in the worst case,
node $i$ cannot respond to the local read query with tag $g_i$ before physical time reaches
\[
R_{ij} = \max(T'_i, \mathcal{T}(g_i) + O_i) = \mathcal{T}(g_i) + \max(\mathcal{L}_{ij}, O_i),
\]
using $T'_i$ given by (\ref{eq:wait2}).
Hence, the \textbf{unavailability} for node $i$ caused by communication from node $j$ is at most
\begin{equation}\label{eq:unavailability}
\bar{A}_{ij} = R_{ij} -  \mathcal{T}(g_i) = \max(\mathcal{L}_{ij}, O_i).
\end{equation}
Combine this with (\ref{eq:unsummed}) and we get the unavailability at process $i$ for a strongly consistent system to be
\begin{eqnarray}\label{eq:unavailability2}
\bar{A}_{i} &=& \max_{j \in N} \max(\mathcal{L}_{ij}, O_i) \\
&=&\max(O_i, \max_{j \in N} \mathcal{L}_{ij}) .
\end{eqnarray}

When network latency, clock synchronization error, or execution time increase sufficiently,
the apparent latency will increase, and so will the unavailability.
We can interpret ``network partitioning'' to mean that $\mathcal{L}_{ij}$ goes to infinity and the system becomes unavailable at process $i$, $\bar{A}_{i} = \infty$.
When enforcing strong consistency, therefore, network partitioning implies unavailability, as expected from the CAP theorem.

\subsection{The CAL Theorem for Arbitrary Consistency}

We can now explore what happens if we relax consistency.
An extreme solution is to eliminate timestamps altogether and have each node just handle messages in the order received.
However, the breach of causal consistency in Fig.~\ref{fig:observer} at $d_2$ becomes hard to avoid,
and assigning an order to distributed user inputs becomes impossible.

A more controlled way to relax consistency is to explicitly manipulate the tags.
In Section~\ref{sec:example}, we will describe a language that supports such manipulation, but for now,
assume that each sending of any update message from node $j$ to $i$ increments the timestamp part of the tag of the message by some fixed amount $D_{ij}$
called a \textbf{logical delay} from $j$ to $i$.
That is, if the send event at $j$ has tag $g_j$ and the corresponding receive event at node $i$ has tag $g_i$, then
\begin{equation}\label{eq:delay2}
\mathcal{T}(g_i) = \mathcal{T}(g_j) + D_{ij}.
\end{equation}
If there is no communication from $j$ to $i$, then we take $D_{ij} = \infty$.
Solving for $D_{ij}$, we get
\begin{equation}\label{eq:delay}
D_{ij} = \mathcal{T}(g_i) - \mathcal{T}(g_j).
\end{equation}

If each node processes events in tag order, and if the merge operation is commutative and associative for updates with identical tags,
then $D_{ij} = \bar{C}_{ij}$, the inconsistency from Definition~\ref{def:consistency}.
Hence, by controlling the logical delay $D_{ij}$, we control the amount of inconsistency between nodes $i$ and $j$.
In exchange for increasing the inconsistency, we potentially reduce the unavailability by $\bar{C}_{ij}$.
Equation (\ref{eq:unavailability}) is replaced by
\begin{equation}\label{eq:unavailability3}
\bar{A}_{ij} = \max(\mathcal{L}_{ij} - \bar{C}_{ij}, O_i).
\end{equation}
This represents the unavailability at $i$ due to messages it may receive from $j$.
One way to understand this is to realize that the $\mathcal{L}_{ij}$ in Definition~\ref{def:apparent} is the difference between
the physical time an update message is received at $i$ and the tag of the update at $j$.
But now, the tag seen at $i$ has been incremented by $D_{ij}$, so if it were to compare the
physical time against the \emph{received} tag (instead of the sender's tag), the latency is reduced by $D_{ij}$.
$D_{ij}$ equals $\bar{C}_{ij}$ because the \emph{same update} has two different logical times on the two nodes
and hence $D_{ij}$ measures the inconsistency these two nodes experience.

The overall unavailability for node $i$ is therefore
\begin{eqnarray}
\bar{A}_i & = & \max_j \left ( \max(\mathcal{L}_{ij} - \bar{C}_{ij}, O_i) \right ) \nonumber \\
& = & \max \left ( O_i, \max_{j \in N} (\mathcal{L}_{ij}- \bar{C}_{ij}) \right ) .
\end{eqnarray}
We have derived the \textbf{CAL theorem} (or $\bar{C}\bar{A}\mathcal{L}$ theorem), which we now state formally:
\begin{theorem}\label{thm:cal}
Given a trace as defined in Section~\ref{sec:inconsistency}, the unavailability in Definition~\ref{def:availability} at process $i$ is, in the worst case,
\begin{equation}\label{eq:cal}
\bar{A}_i = \max \left ( O_i, \max_{j \in N} (\mathcal{L}_{ij} - \bar{C}_{ij}) \right ) ,
\end{equation}
where $O_i$ is the processing offset given by Definition~\ref{def:processing},
$\mathcal{L}_{ij}$ is apparent latency in Definition~\ref{def:apparent} (which includes $O_j$), and
$\bar{C}_{ij}$ is the inconsistency of Definition~\ref{def:consistency} .
\end{theorem}
Note from Definition~\ref{def:consistency} that if there is a write event on $j$ 
and no corresponding receive event on $i$, then the inconsistency becomes $\bar{C}_{ij} = \infty$.
This corresponds to the case of total network partitioning.
When this is the case, however, (\ref{eq:cal}) tells us that node $j$ has no effect on availability on node $i$, as expected.

\subsection{More on Processing Offsets}

The processing offsets $O_i$ and $O_j$ deserve more discussion.
These are physical time delays incurred on nodes $i$ and $j$ before they can begin handling events with a particular tag.
Specifically, node $i$ can begin handling a user input with tag $g_i$ at physical time $T_i = \mathcal{T}(g_i) + O_i$.
We will see that these offsets are derivable from program structure, but for this simple bulletin board application,
as we will see in Section~\ref{sec:implementation},
they can both be made zero, $O_i = O_j = 0$. 
This means that each node can immediately launch local posts into the network, incurring only
an execution time delay $X_{ij}$.
This will not compromise causal consistency in any way.

For strong consistency, we require $\bar{C}_{ij} = 0$.
In this case, the time it takes to respond
to a user on node $i$ is the largest apparent latency from another node $j$.
If we are willing to tolerate an inconsistency at least as large as this apparent latency,
i.e., if
\begin{equation}\label{eq:threshold}
\bar{C}_{ij} \ge  \mathcal{L}_{ij} \text{ for all nodes } j,
\end{equation}
then the unavailability becomes zero for the case where the processing offsets are zero.
We can respond immediately to user input.
This is a reasonable design choice for a bulletin board application because users will expect to see their posts immediately, as they type them,
and given timestamped posts, the record of the conversation is easily repaired when other posts are later received.
If network latency increases so that (\ref{eq:threshold}) no longer holds, then we can either tolerate larger
inconsistency, allowing $\bar{C}_{ij}$ to rise, or tolerate higher unavailability, allowing $\bar{A}_i$ to rise.
For the bulletin board application, the former is probably the better choice under the assumption that the network will eventually be repaired.

The CAL theorem gives a relation between inconsistency, unavailability, and apparent latency,
where causal consistency is ensured at the price of
processing offsets $O_i$. We will show in Section~\ref{sec:implementation} how to derive these offsets in \lf from program structure.

\subsection{Ensuring Eventual Consistency}

To achieve eventual consistency, if $\bar{C}_{ij} > 0$, we have a bit more work to do.
To improve availability, we introduce logical delays $D_{ij}$, which then determine the inconsistency $\bar{C}_{ij}$.
To achieve eventual consistency, all processes must eventually behave as if they had applied writes in the same order,
but the logical delays permit them to apply the writes in different orders.
For the bulletin board application, we can fix this by modifying the merge operation to use the
\emph{original} tag, before incrementing by the logical delay $D_{ij}$, to perform the merge.
That is, if a node $i$ receives a message from node $j$ with timestamp $t_j$, it knows that the original timestamp was $t_j - D_{ij}$.
The receiving end can use this original tag to sort updates, thereby ensuring that all replicas eventually
have the same sequence of posts to the bulletin board.
We call this style of merge operation a \textbf{sorted append}.

When the merge operation is \textbf{replace}, we can similarly modify the merge to become a \textbf{sorted replace}
which becomes associative and commutative.
At the receiving end, we can use (\ref{eq:delay2}) to compute the original timestamp $\mathcal{T}(g_j) = \mathcal{T}(g_i) - D_{ij}$.
However, we need a bit more, because the order of the updates is determined by their \emph{tags}, not their timestamps.
Hence, it will be necessary to either convey the original tag in a message or ensure that there is a way to calculate it at the receiving end
(for example, in \lf, by conveying the microstep).
We can compare this original tag against the tag of the latest local update to determine whether the received value should
overwrite the previously applied local update.
This strategy enforces a \textbf{last writer wins} policy in a rigorous, well-defined way.

\subsection{Max-Plus Formulation}

Summarizing where we are, the CAL theorem gives an algebraic relation between inconsistency, unavailability, and latency.  
This relation shows that if network latency becomes unbounded (the network becomes partitioned), 
then one of inconsistency and unavailability must also become unbounded, and hence the CAP theorem is a special case of the CAL theorem. 
The relation involves processing offsets, representing the cost of enforcing causal consistency.
How to derive these is somewhat implementation dependent.
In Section~\ref{sec:example}, we show how to use the \lf coordination language to specify such
distributed programs in a way that enables calculation of the offsets,
and in Section~\ref{sec:implementation}, we describe two specific distributed implementations
that we have realized in \lf.
But first, we give a very general form of the CAL theorem.

%
%

The form of the CAL theorem given in (\ref{eq:cal}) can be made much more elegant by observing that
the mathematical operations of maximization and addition on numbers form an algebra called a max-plus
algebra with very useful properties~\cite{Baccelli:92:MaxPlus}.
In this algebra, addition (written $\oplus$) is maximization,
\[
a \oplus b = \max(a,b) ,
\]
and multiplication (written $\otimes$) is addition,
\[
a \otimes b = a + b .
\]
In max-plus, the additive identity is $-\infty$,
and the multiplicative identify is zero.

Let $N$ be the number of nodes, and define an $N \times N$ matrix $\Gamma$ such that its elements are given by
\begin{equation}\label{eq:matrix}
\Gamma_{ij} = \mathcal{L}_{ij} - \bar{C}_{ij} - O_j .
\end{equation}
That is, from (\ref{eq:latency}), the $i$, $j$-th entry in the matrix is an assumed bound on 
$X_{ij} + L_{ij} + E_{ij}$ (execution time, network latency, and clock synchronization error),
adjusted downwards by the specified tolerance for inconsistency.

Let $\bm{A}$ be a column vector with elements equal to the unavailabilities $\bar{A}_i$,
and $\bm{O}$ be a column vector with elements equal to the processing offsets $O_i$.
Then the CAL theorem (\ref{eq:cal}) can be written as
\begin{equation}\label{eq:calmatrix}
\bm{A} = \bm{O} \oplus \Gamma \bm{O},
\end{equation}
where the matrix multiplication is in the max-plus algebra.
This can be rewritten as
\begin{equation}\label{eq:calmatrix2}
\bm{A} = (\bm{I} \oplus \Gamma) \bm{O},
\end{equation}
where $\bm{I}$ is the identity matrix in max-plus, which has zeros along the diagonal and $-\infty$ everywhere else.
Hence, unavailability is a simple linear function of the processing offsets,
where the function is given by a matrix
that depends on the network latencies, clock synchronization error, execution times, and specified inconsistency in a simple way.

When the processing offsets are zero, as they are for the bulletin board application,
then the unavailability for each node is the sum (in max-plus) of the corresponding row of $\bm{I} \oplus \Gamma$ which is either zero
(if all elements in the row of $\Gamma$ are less than or equal to zero) or the maximum element in the row of $\Gamma$.

\subsection{Pessimistic Evaluation of Processing Offsets}\label{sec:pessimistic}

The processing offsets $O_i$ and $O_j$ are physical time delays incurred on nodes $i$ and $j$ before they can begin handling events.
Specifically, node $i$ can begin handling a user input (specifically a write event) with tag $g_i$ at physical time $T_i = \mathcal{T}(g_i) + O_i$.
In the absence of any further information about a program, we can use our $\Gamma$ matrix to calculate these offsets.
However, we will find that the result is pessimistic and can be refined with further information about the program structure.

First, consider nodes that have the possibility of new events appearing asynchronously with timestamps given by the local physical clock,
like those in our bulletin board application.
In such a node $i$, it is generally not safe to process an event with tag $g_i$ until the physical clock $T_i$ exceeds $\mathcal{T}(g_i)$, i.e., $T_i > \mathcal{T}(g_i)$.
Otherwise, there is a possibility of processing events out of order.
If a node $j$ is purely functional and only reacts to network inputs, then there is no such constraint.
Define a column vector $\bm{Z}$ such that
\begin{equation}\label{eq:Z}
Z_i = \begin{cases}
0, & \text{if node $i$ has local physically timestamped inputs,} \\
-\infty, & \text{otherwise}.
\end{cases}
\end{equation}
With this, we require at least that
\[
\bm{O} \ge \bm{Z}.
\]
In addition, to ensure that node $i$ processes events in tag order, it is sufficient to ensure that node $i$ has received all network input
events with tags less than or equal to $g_i$ before processing any event with tag $g_i$.
With this (conservative) policy,
\[
O_i \ge \max_j (\mathcal{L}_{ij}- \bar{C}_{ij}) .
\]
The smallest processing offsets that satisfy these two constraints satisfies
\begin{equation}\label{eq:pessimistic}
\bm{O} = \bm{Z} \oplus \Gamma \bm{O} .
\end{equation}
This is a system of equations in the max-plus algebra.
From Baccelli, et al. \cite{Baccelli:92:MaxPlus} (Theorem 3.17),
if every cycle of the matrix $\Gamma$ has weight less than zero, then the unique solution of this equation is
\begin{equation}\label{eq:solution}
\bm{O} = \Gamma^* \bm{Z},
\end{equation}
where the \textbf{Kleene star} is (Theorem 3.20 \cite{Baccelli:92:MaxPlus})
\[
\Gamma^* = \bm{I} \oplus \Gamma \oplus \Gamma^2 \oplus \cdots .
\]
Baccelli et al. show that this reduces to
\[
\Gamma^* = \mathbf{I} \oplus \Gamma \oplus  \cdots \oplus \Gamma^{N-1},
\]
where $N$ is the number of processes.

The requirement that the cycle weights be less than zero is intuitive,
but overly restrictive.
It means that along any communication path from a node $i$ back to itself,
the sum of the logical delays $D_{jk}$ must exceed the sum of the execution times, network latencies, and clock synchronization errors
along the path.
This means that we have to tolerate a non-zero inconsistency somewhere on each cycle.
For the bulletin board application, where every node sends messages to every other node,
every pair of nodes requires a non-zero inconsistency in order to satisfy this cycle-mean constraint.

In practice, programs may have zero or positive cycle means.
Theorem 3.17 of Baccelli, et al. \cite{Baccelli:92:MaxPlus} shows that if all cycle weights are non-positive,
then there is a solution, but the solution may not be unique.
If there are cycles with positive cycle weights, there is no finite solution for $\bm{O}$.
When the inconsistency is zero,
there are no logical delays at all, 
and all cycle weights become positive.
In this case, the only solution to (\ref{eq:pessimistic}) sets
all the processing offsets to $\infty$.
Every node must wait forever before handling any user input.
This is, of course, the ultimate price in availability.

Why is (\ref{eq:solution}) pessimistic?
Absent further information about the application logic, we must assume that any network input at node $i$
with tag $g_i$ can causally affect any network output with tag $g_i$ or larger.
Moreover, we have no knowledge about where network inputs may originate.
We next show how, by constructing the application in a language that exposes more
information about causal relationships, we can derive much less pessimistic processing offsets
while still preserving causal and eventual consistency.

\section{Lingua Franca Realization}\label{sec:example}

\lf (or \lfshort, for short) is a coordination language developed jointly at UC Berkeley, TU Dresden, UT Dallas, and Kiel University~\cite{LohstrohEtAl:21:Towards}.
We show here that \lfshort supports a full range of
explicit tradeoffs between availability and consistency
for a wide variety of applications.
\lf is a polyglot coordination language that orchestrates
concurrent and distributed programs written in any of several target languages
(as of this writing, C, C++, Python, and TypeScript are supported, and Rust is in progress).

In this section, we give a small collection of complete \lf programs that illustrate these tradeoffs.\footnote{Download these programs from \url{https://cal.lf-lang.org/}.}
To keep the programs short but complete, we develop an example borrowed from Kuhn \cite{Kuhn:17:Reactive}
that has similar properties to the bulletin board considered earlier, but does not require memory management for a growing text
and hence results in simpler code that fully illustrates the principles of interest.

Kuhn's application maintains bank balances in a distributed database
and accepts deposits and withdrawals at distributed locations like ATM machines.
Suppose that a customer shows up and wants to withdraw $w$ dollars.
This needs to be compared against the current balance $x$ in the account.
However, there may be near-simultaneous withdrawals occurring at other locations.
Strong consistency requires agreement on the order in which these withdrawals occur.
If all locations agree on this order, the bank can assure that the balance never drops below zero by
denying any withdrawal that would make it so.
Moreover, if the bank policy allows the balance to drop below zero, all branches will agree on the number of
times that this has occurred, so they will all agree on what overdraft charges to apply.

When network latency gets large, however, such a strong consistency policy reduces availability by making the user wait.
In the extreme case of network partitioning, the ATM may simply deny the service rather than wait for the network to be repaired.
We can define ``network partitioning'' to be ``sufficiently large network latency upon which communication retries are abandoned.''
The threshold of network partitioning, therefore, like the tradeoff between availability and consistency, is defined by the application.

Kuhn's banking example is a good one because we can image a whole range of design choices.
Ultimately, the choices are business decisions, not technical ones.
Is it acceptable for an ATM to deny dispensing cash because of a temporary network failure?
How would customers react?
How big is the risk to the bank if the ATM dispenses cash on the basis of possibly inconsistent data?
Banks routinely allow accounts to go into overdraft, and then heap fees on their customers.
Then again, a distributed attack that simultaneously makes a large number of near-simultaneous withdrawals after launching
a distributed denial of service attack on a bank's networks is not hard to imagine.
The tradeoff between risk and customer service is a business decision.
We will show software tools that make these choices clear and enable a whole range of solutions.

This banking example also has the nice property that the database operations are associative and commutative
if overdrafts are allowed, and not otherwise.
Hence, we can explore designs with a variety of merge operations (associative, commutative, both, or neither).

\subsection{Brief Introduction to Lingua Franca}

\begin{figure}[tb]
\begin{lstlisting}[style=framed,language=LF,escapechar=|]
target L;
reactor ReactorClass {
    input name:type;    |\label{ln:input}|
    ...
    output name:type;    |\label{ln:output}|
    ...
    state name:type(init);   |\label{ln:state}|
    ...
    ... timers, actions, if any ...
    ...
    reaction(trigger, ...) -> effect, ... {=  |\label{ln:reaction}|
        ... code in language L ...
    =}
    ... more reactions ...
}
...
federated reactor {
    instance = new ReactorClass(); |\label{ln:instance}|
    ...
    instance.name -> instance.name; |\label{ln:connection}|
    ...
}
\end{lstlisting}
\caption{Structure of a federated \lf program for target language $L$. \label{fig:lf}}
\end{figure}

\lf\footnote{\url{https://repo.lf-lang.org}} is a coordination language where applications are defined as concurrent compositions of components called
\textbf{reactors}~\cite{Lohstroh:2019:CyPhy,Lohstroh:EECS-2020-235}.
Fig.~\ref{fig:lf} outlines the structure of a \lf program.
One or more \textbf{reactor classes} are defined with \textbf{input ports} (line \ref{ln:input}),
\textbf{output ports} (line \ref{ln:output}),
\textbf{state variables} (line \ref{ln:state}),
and timers and actions.
We will not need timers here and will elaborate on actions later.
If a reactor class is instantiated within a \textbf{federation}, as shown on line \ref{ln:instance},
then the instance is called a \textbf{federate}, and
tagged inputs will arrive from the network at the input ports and be handled in tag order.
Inputs are handled by \textbf{reactions}, as shown on line \ref{ln:reaction}.
Reactions declare their \textbf{triggers}, as on line \ref{ln:reaction}, which can be input ports, timers, or actions.
If a reaction lists an output port among its \textbf{effects}, then it can produce tagged output messages
via that output port.
The routing of messages is specified by \textbf{connections}, as shown on line \ref{ln:connection}.
The syntax and semantics will become clearer as we develop our specific applications.

\begin{figure}[h!]
    \begin{minipage}{0.70\linewidth} 
\begin{lstlisting}[style=framed,language=LF,escapechar=|]
target C;
reactor CAReplica {
    input local_update:int;    |\label{ln:replica_in_1}|
    input remote_update:int;
    input query:bool;       |\label{ln:replica_in_n}|
    
    state balance:int(0);   |\label{ln:balance}|

    output response:int;    |\label{ln:replica_out}|
    
    reaction(local_update, remote_update) {=  |\label{ln:replica_reaction1}|
        if (local_update->is_present) {
            self->balance += local_update->value;
        }
        if (remote_update->is_present) {
            self->balance += remote_update->value;
        }
    =}    |\label{ln:replica_reactionn}|
    
    reaction(query) -> response {= |\label{ln:replica_response1}|
        SET(response, self->balance);
    =}|\label{ln:replica_responsen}|
}
\end{lstlisting}
\end{minipage}
\begin{minipage}{0.29\linewidth}
\includegraphics[width=\linewidth]{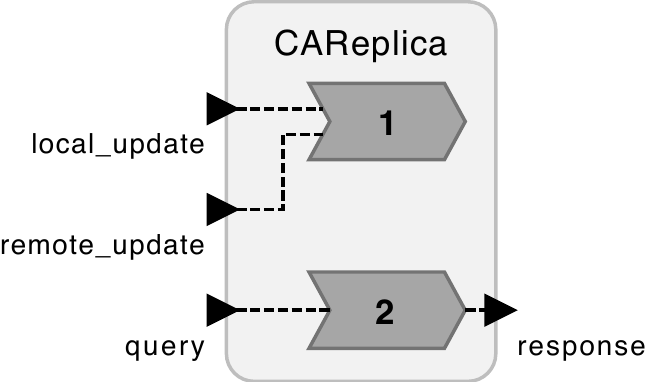}
\end{minipage}
\caption{\lf code defining a reactor class that is a replica in a replicated database storing a bank balance and accepting queries and updates. This version is commutative and associative. \label{fig:replica}}
\end{figure}

\subsection{Commutative and Associative Replica}
We begin with a version that has an associative and commutative merge operation.
The first part of this version is shown in Fig~\ref{fig:replica}.
This defines a reactor class, a software component that is described as an ``actor revisited''~\cite{DBLP:conf/dac/LohstrohSGWGSL19}.
The first line defines the \textbf{target language}, which is the language of the program that the \lfshort
code generator will produce and the language in which the business logic of the software component is written.
To minimize dependencies, we give our examples here with C as the target language.

The second line declares a new reactor class called \texttt{CAReactor} (`CA' for Commutative and Associative).
This reactor has three inputs, defined on lines \ref{ln:replica_in_1} through \ref{ln:replica_in_n}.
The first input accepts a local update, an integer that is positive for a deposit and negative for a withdrawal.
The second input accepts a remote update, which will come from some other machine somewhere on the network
(we will generalize this later to accept an arbitrary number of remote updates).
The third input accepts a query for the current balance.

Line \ref{ln:balance} defines a local state variable, an integer that is the local copy of the balance.
Line \ref{ln:replica_out} defines an integer output, which will be a response to a query for the
balance. 
In a strongly consistent design (which we will see how to construct), this balance will be agreed upon by all replicas at each tag.

Lines \ref{ln:replica_reaction1} through \ref{ln:replica_reactionn} give the business logic
of how to handle local or remote updates. The code between the delimiters \texttt{\{= ... =\}} is ordinary C code
making use of mechanisms provided by the \lfshort code generator to access the inputs and state.
This code simply checks to see which inputs are present and adjusts the balance accordingly.
Because the operation is commutative and associative, it does not matter in which order these inputs are handled.

The figure to the right is automatically generated by the \lf IDE called Epoch.\footnote{Epoch is available for download at \url{https://releases.lf-lang.org}.}\textsuperscript{,}\footnote{The diagram synthesis feature was created
by Alexander Schulz-Rosengarten of Kiel University using the graphical layout tools from the
KIELER Lightweight
Diagrams framework~\cite{SchneiderSvH13} (see \url{https://rtsys.informatik.uni-kiel.de/kieler}).}
The chevrons in the figure represent reactions, and their dependencies on inputs and their ability to produce outputs
is shown using dashed lines.


\subsection{User}
\begin{figure}[tb]
    \begin{minipage}{0.65\linewidth} 
\begin{lstlisting}[style=framed,language=LF,escapechar=|]
target C;
reactor UserInput {
   preamble {=
      // Define a function to read user input.
      void* get(void* r) {  |\label{ln:thread1}|
         int amt;
         char buf[20];
         while(1) {
            // Read a character input.
            char* ln = fgets(buf, 20, stdin);|\label{ln:block}|
            // Exit if no more input.
            if (ln == NULL) return NULL;
            // Parse an input integer.
            int n = sscanf(ln, "%d", &amt);
            // If one integer was parsed...
            if (n == 1) {
               // Schedule an event.
               schedule_int(r, 0, amt);|\label{ln:schedule}|
            } else {
               // Request another input.
               printf("Please enter a number.\n");
            }
         }
      }   |\label{ln:threadn}|
   =}
   input balance:int;
   output deposit:int;
   
   physical action r:int;|\label{ln:physicalaction}|
   
   reaction(startup) -> r {=
      pthread_t id;
      pthread_create(&id, NULL, &get, r);|\label{ln:threadstart}|
   =}
   reaction(r) -> deposit {=|\label{ln:deposit1}|
      SET(deposit, r->value);
   =}|\label{ln:depositn}|
   reaction(balance) {=|\label{ln:reactionbalance}|
      printf("Balance: %d\n", balance->value);
   =}|\label{ln:reactionbalancen}|
}
\end{lstlisting}
\end{minipage}
\begin{minipage}{0.34\linewidth}
\includegraphics[width=\linewidth]{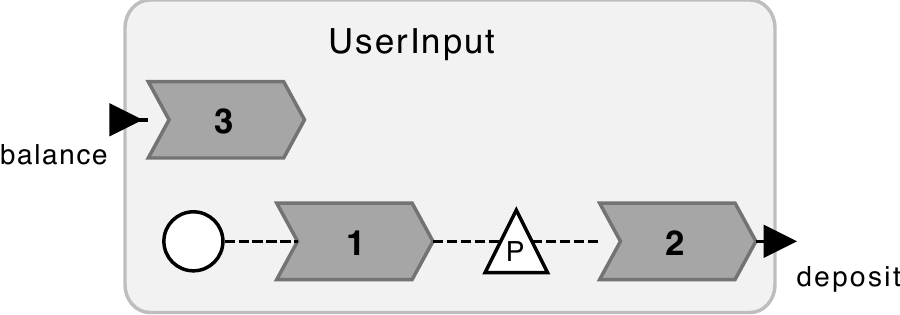}
\end{minipage}
\caption{\lf component that gets user input to provide deposits and make withdrawals. \label{fig:user}}
\end{figure}

The code in Fig.~\ref{fig:user} defines a \lf reactor that stands in for an ATM machine through which a customer can make
deposits or withdrawals.
This component listens for the user to type a number on a terminal, and then produces that number on its output port.
To keep things simple, there is no authentication and not much error checking. While obviously critical to a real ATM design, those 
aspects are irrelevant to our discussion here, so we leave them out.


Upon startup, on line \ref{ln:threadstart}, this reactor creates a thread that executes concurrently with the \lf program.
That thread, defined on lines \ref{ln:thread1} to \ref{ln:threadn}, repeatedly blocks on line \ref{ln:block} waiting for the user to type something.
If the user input is a valid number, then on line \ref{ln:schedule}, the thread calls a built-in thread-safe function \texttt{schedule},
passing it a pointer to the \textbf{physical action} named `\texttt{r}' and the amount entered by the user (the 0 argument is irrelevant
to the current discussion).

The \textbf{physical action} \texttt{r}, declared on line  \ref{ln:physicalaction}, is a \lf construct for providing
external, asynchronous inputs to an \lfshort program.
The key is that when \texttt{schedule} is called, a tag $g$ based on a local measurement $T$ of physical time,
such that $\mathcal{T}(g) = T$,
is assigned to the event, which is then injected into the program to be handled in tag order.

The \lfshort program reacts to the event created by the call to \texttt{schedule} by executing the reaction given on lines \ref{ln:deposit1} through \ref{ln:depositn}.
This sets the output named \texttt{deposit} to the amount entered by the user.

The \textbf{physical action} declared on line  \ref{ln:physicalaction}
of Fig.~\ref{fig:user} deserves more scrutiny.
First, \lf, by default, uses the system clock on the machine that runs each federate to assign the timestamp part $\mathcal{T}(g)$ of the tag.
When a federated program is started, it performs a clock synchronization round using the technique of Geng et al. \cite{GengEtAl:18:ClockSync}.
This ensures that even if the system clock is set manually to some arbitrary value, when the distributed program starts up,
all nodes will agree on the current physical time within a few milliseconds.
\lf also provides a facility for performing ongoing clock synchronization that can correct for clock drifts, but in many systems,
it is not really necessary to enable this.
We can rely instead on a built-in NTP realization, if that is sufficiently precise for the application.

Once a tag is assigned, the handling of the update throughout the distributed system is a deterministic function
of that tag. This gives a clear semantics to the behavior of the system when the actual order of events
originating throughout the system is unknown, unknowable,
or ambiguous. Moreover, it enables rigorous regression testing, where the \texttt{UserInput} reactor of Fig.~\ref{fig:user} is replaced
by event generators driven by logical clocks. Those logical clocks can generate events on distributed nodes that are deterministically ordered
or even simultaneous.

\subsection{Composition}

\begin{figure}[tb]
\begin{lstlisting}[style=framed,language=LF,escapechar=|]
target C;
import CAReplica from "CAReplica.lf";
import UserInput from "UserInput.lf";

reactor ATM {
    input update:int;
    output publish:int;
    u = new UserInput();
    r = new CAReplica();
    (u.deposit)+ -> r.query, r.local_update;|\label{ln:multi}|
    r.response -> u.balance;
    update -> r.remote_update;
    reaction(u.deposit) -> publish {=|\label{ln:filter1}|
        if (u.deposit->value != 0) {
            SET(publish, u.deposit->value);
        }
    =}|\label{ln:filtern}|
}
federated reactor ConsistencyFirst {
    a = new ATM();
    b = new ATM();
    b.publish -> a.update;
    a.publish -> b.update;
}
\end{lstlisting}
\includegraphics[width=\linewidth]{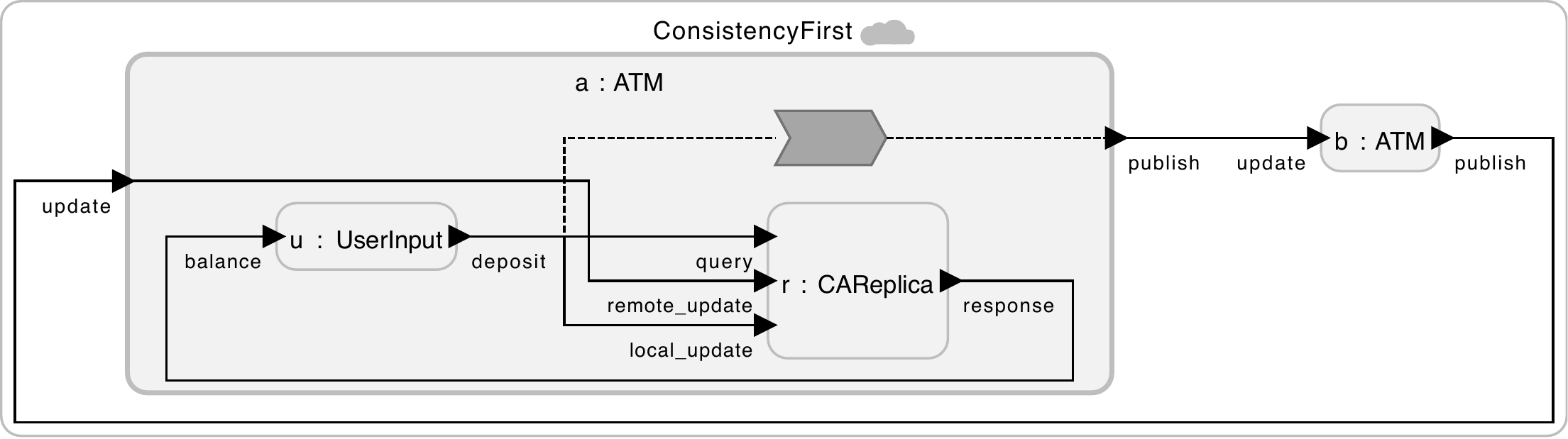}
\caption{Federated \lf program with two ATM machines that can provide deposits and make withdrawals. \label{fig:consistencyfirst}}
\end{figure}

We can now put together the components to get a complete, executable program.
In Fig~\ref{fig:consistencyfirst}, we define an \texttt{ATM} reactor that contains one instance each of
\texttt{UserInput} from Fig.~\ref{fig:user} and \texttt{CAReplica} from Fig.~\ref{fig:replica}.
We then define a \textbf{federated} reactor named \texttt{ConsistencyFirst}
that creates two instances of \texttt{ATM} and connects them.
Hopefully, the program, with the help of the diagram, is self-explanatory, with the possible exception of the new syntax on line \ref{ln:multi}.
To the left of the arrow, the output port \texttt{u.deposit} is surrounded with \texttt{( ... )+}, which, in \lf syntax,
indicates to use the port as many times as necessary to satisfy all the destinations given
to the right of the arrow. In other words, it is a compact syntax for multicast.
The reaction defined on lines \ref{ln:filter1} through \ref{ln:filtern} simply prevents publishing zero-valued deposits.
Hence, a deposit equal to zero can be used to query the current balance and will not generate network traffic.

\subsection{Execution}


When the top-level reactor in a \lf program is \textbf{federated}, as it is in Fig.~\ref{fig:consistencyfirst}, then the code generator,
instead of producing a single program, produces as many programs as there are instances of reactors within the top level reactor,
plus one additional program that manages certain coordination functions (more about this later).
In this case, there are two reactors within the top level, so a total of three programs will be generated.

\begin{figure}[tb]
\centering
\includegraphics[width=0.7\columnwidth]{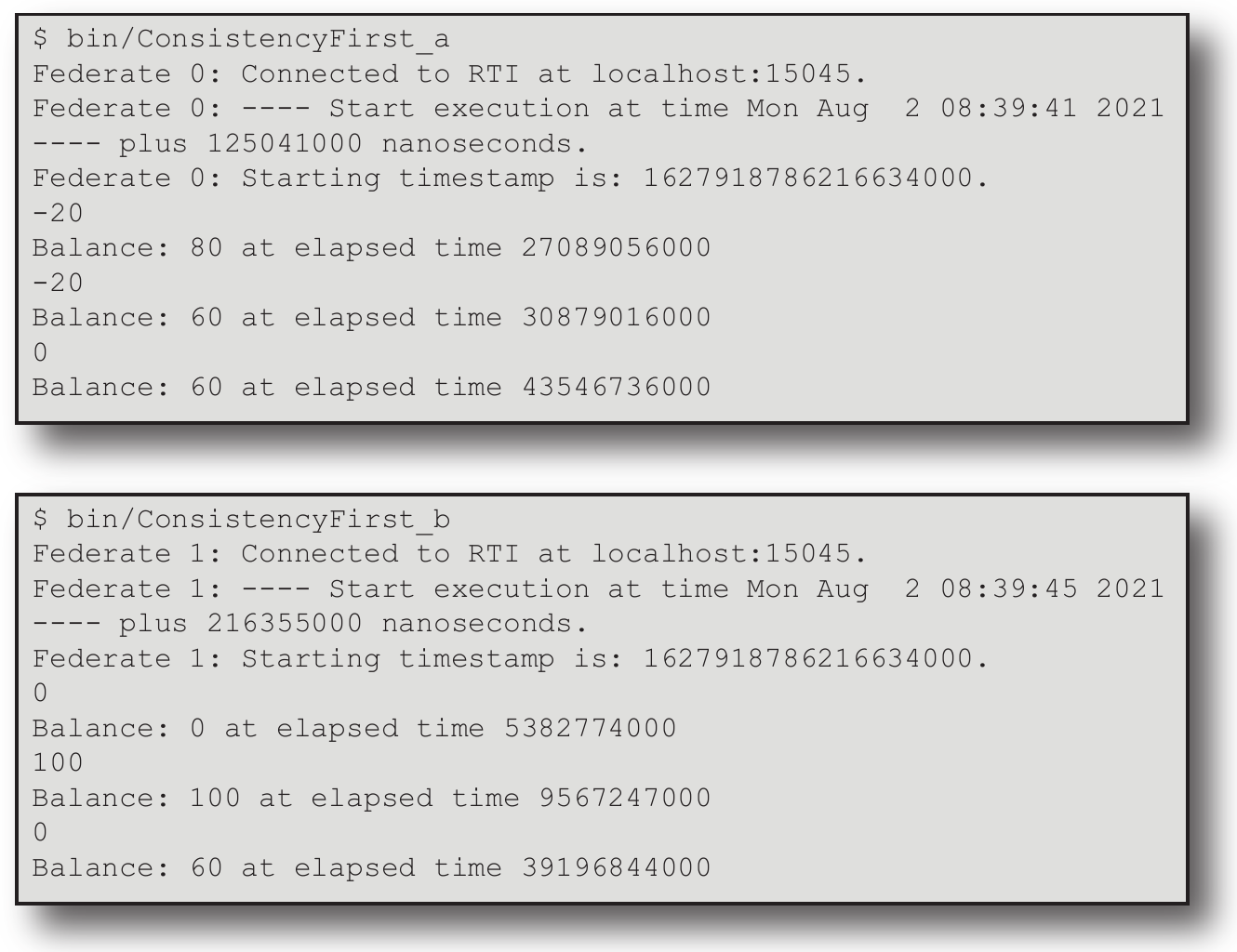}
\caption{An execution of the \lf program in Fig.~\ref{fig:consistencyfirst} (slightly elaborated to display (elapsed) logical time $\mathcal{T}(g)$ (in nanoseconds). \label{fig:execution}}
\end{figure}

An execution of the program in Fig.~\ref{fig:consistencyfirst} is shown in Fig.~\ref{fig:execution},
where there is one terminal for each of the two \texttt{ATM} instances.
The top line of each window shows the command that starts each instance.
In the lower window, user $b$ begins by querying the current balance by entering `0'.
The balance is zero.
User $b$ then deposits 100 dollars.
User $a$ then withdraws 20 dollars twice.
User $b$ then queries the balance again, discovering that it is now 60.

The deposits and withdrawals are handled by both \texttt{ATM} federates in the same order,
defined by tags that are assigned when \texttt{schedule} is called on line \ref{ln:schedule} in Fig.~\ref{fig:user}.
If two deposits occur with the same tag, then the reported balance by each user will reflect
the aggregate of the two operations.
That is, the two deposits are semantically \textbf{simultaneous}.

This simultaneity feature is hard to test with an interactive program like this,
but in \lf, it is easy to create a regression test that replaces the \texttt{UserInput} with
timer-driven inputs, which gives precise control over the tags.
The ability to construct such deterministic, distributed regression tests is one of the key
advantages of \lf.

\subsection{Variants of the Running Example}

Lest the reader conclude that we are only talking about one rather oversimplified example,
we will now point out several variants of this design that are easy to build.

\begin{figure}[tbp]
\begin{lstlisting}[style=framed,language=LF,escapechar=|]
target C;
import UserInput from "UserInput.lf";

federated reactor ConsistencyFirstN(
    N:int(4)|\label{ln:param}|
) {
    bank = new[N] ATM(N = N);|\label{ln:bank}|
    (bank.publish)+ -> bank.updates;|\label{ln:bankconnection}|
}
reactor ATM(
    N:int(2)|\label{ln:param2}|
) {
    input[N] updates:int;
    output publish:int;
    u = new UserInput();
    r = new CAReplicaN(N = N);
    u.deposit -> r.query;
    r.response -> u.balance;
    updates -> r.updates;
    reaction(u.deposit) -> publish {=
        if (u.deposit->value != 0) {
            SET(publish, u.deposit->value);
        }
    =}
}
reactor CAReplicaN(
    N:int(2)
) {
    input[N] updates:int;|\label{ln:multiport}|
    input query:int;
    output response:int;
    state balance:int(0);
    
    reaction(updates) {=|\label{ln:iterate}|
        for (int i = 0; i < updates_width; i++) {
            if (updates[i]->is_present) {
            	self->balance += updates[i]->value;
            }
        }
    =}|\label{ln:iteraten}|
    reaction(query) -> response {=|\label{ln:queryN}|
        SET(response, self->balance);
    =}
}
\end{lstlisting}
\includegraphics[width=\linewidth]{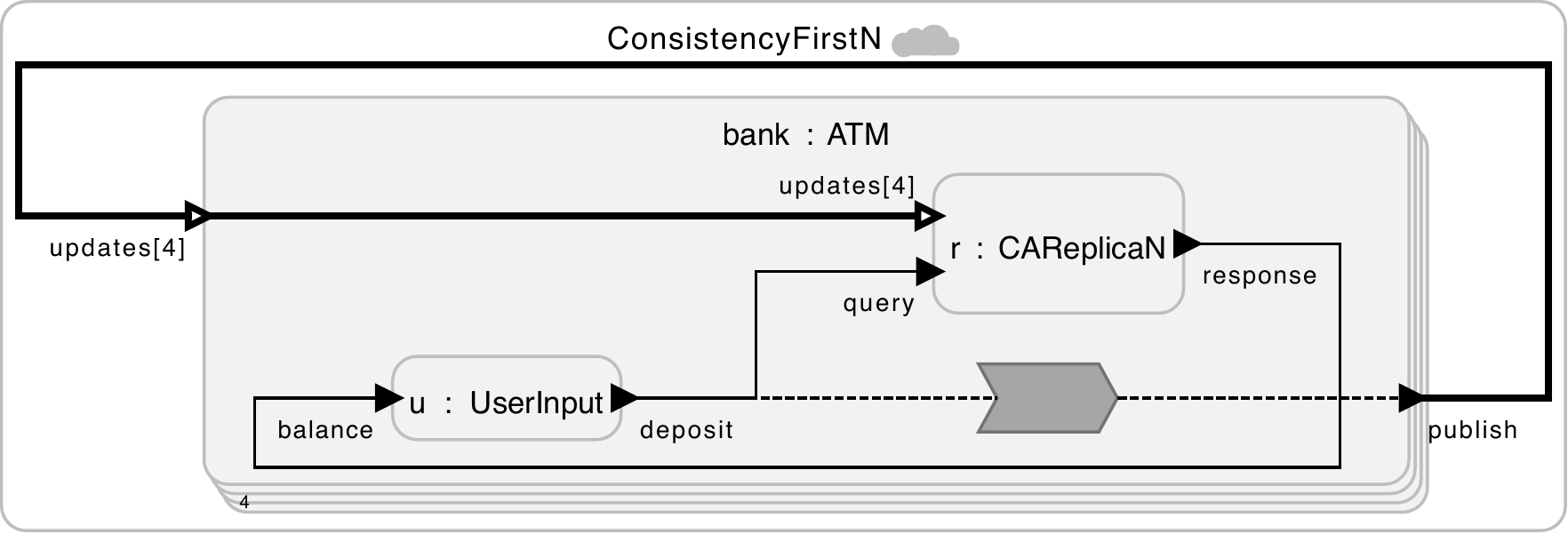}
\caption{Federated \lf program with any number of ATM machines that can provide deposits and make withdrawals. \label{fig:consistencyfirstn}}
\end{figure}

First, the example in Fig.~\ref{fig:consistencyfirst} has only two federates.
We have extended
\lf with a convenient syntax 
shown in Fig.~\ref{fig:consistencyfirstn} for scaling this program to any number of federates.
The top-level \texttt{ConsistencyFirstN} reactor has a parameter \texttt{N} defined on line \ref{ln:param} (with default value 4) that specifies
how many instances of the ATM reactor to create.
Those instances are created on line \ref{ln:bank}, and each instance is assigned the value \texttt{N} to its own parameter,
defined on line \ref{ln:param2}, which happens to also have the name ``\texttt{N}.''

The \texttt{CAReplicaN} reactor is similar to \texttt{CAReplica} in Fig.~\ref{fig:replica},
with the only difference being its \textbf{multiport} input, defined on line \ref{ln:multiport},
which can accept \texttt{N} input connections.
The reaction on lines \ref{ln:iterate} through \ref{ln:iteraten} iterates over these inputs
and adds to the balance any values it finds.

For the example in Fig.~\ref{fig:consistencyfirstn}, it does not matter in what order simultaneous updates are applied
because the updates are commutative and associative and no replica reads the result until all updates have been applied.
Many distributed applications with shared data, however, do not naturally have commutative and associative merge operations.

\begin{figure}[tbp]
\begin{lstlisting}[style=framed,language=LF,escapechar=|]
target C;
import UserInput from "UserInput2.lf";

federated reactor ReplicatedDataStore(N:int(4)) {
    nodes = new[N] Node(N = N);|\label{ln:banknodes}|
    (nodes.publish)+ -> nodes.updates;|\label{ln:pubconnection}|
}
reactor Node(N:int(2)) {
    input[N] updates:int;
    output publish:int;
    u = new UserInput();
    r = new ReplicaN(N = N);
    u.update -> r.query;
    u.update -> publish;
    r.response -> u.current_value;
    updates -> r.updates;
}
reactor ReplicaN(N:int(2)) {
    input[N] updates:int;|\label{ln:replicamultiport}|
    input query:int;
    output response:int;
    state record:int(0);
    
    reaction(updates) {=
        for (int i = 0; i < updates_width; i++) {|\label{ln:iteration}|
            if (updates[i]->is_present) {
                // Overwrite simultaneous updates on lesser input channels.
            	self->record = updates[i]->value;|\label{ln:overwrite}|
            }
        }
    =}
    reaction(query) -> response {=
        SET(response, self->record);
    =}
}
\end{lstlisting}
\includegraphics[width=\linewidth]{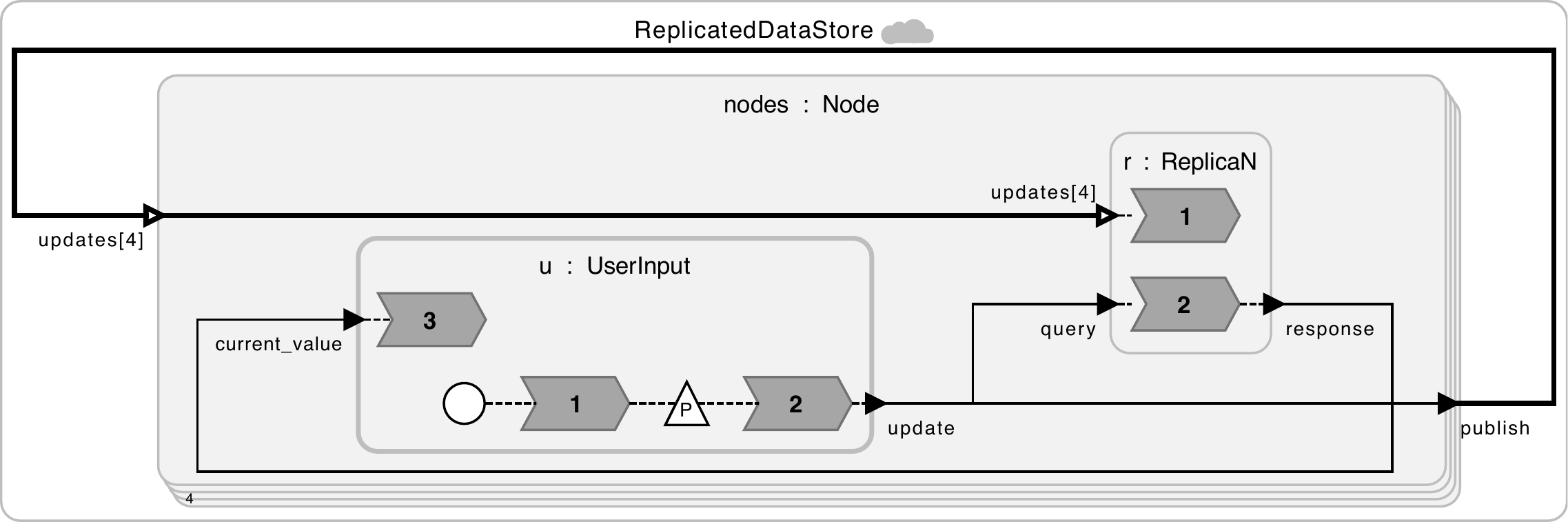}
\caption{Distributed \lf program with any number of nodes that can update a shared data value. \label{fig:replicated}}
\end{figure}

Fig.~\ref{fig:replicated} shows a variant where, on line \ref{ln:overwrite}, an update \emph{overwrites} the shared value.
Such an operation is associative but not commutative.
Here, each user update is broadcast to all nodes and, as before, applied before any query for the value is processed.
If two updates are logically simultaneous, then both updates will appear \emph{in deterministic order} at the
multiport input (line \ref{ln:replicamultiport}) of the replica instance.
Because the order in which these updates appear is deterministic, a priority scheme can be used to determine
which update prevails.
In this implementation, a bank of $N$ nodes is created on line \ref{ln:banknodes},
and each node is assigned a unique \texttt{bank\_index} parameter ranging from 0 to $N-1$,
where $N$ is the \texttt{updates\_width} variable referenced on line \ref{ln:iteration}.
Nodes with higher bank indices have priority over nodes with lower ones simply because
the iteration on line \ref{ln:iteration} reads inputs from other members of the bank in the same
order as their  \texttt{bank\_index}.
Any write by a node with a higher \texttt{bank\_index} will overwrite
a write by a node with a lower \texttt{bank\_index} that is simultaneous.
Hence, simultaneous updates yield deterministic results
prioritized by bank index.

\subsection{Trading Off Consistency and Availability in Lingua Franca}

Unavailability is a measure of the time it takes for a system to respond to user requests.
If it takes a long time (or it never responds), then the system is unavailable, whereas if responses
are instantaneous, then the system is highly available.

In the ATM application in Fig.~\ref{fig:consistencyfirstn},
a user request at the $i$-th ATM is assigned a tag $g_i$ on line \ref{ln:schedule} of Fig.~\ref{fig:user} based on the local physical clock.
Hence, $\mathcal{T}(g_i)$ is a good measure of the physical time at which the user has initiated a request.
The user's request turns into a tagged \texttt{deposit} output from the \texttt{UserInput} reactor, which gets sent to the
\texttt{query} input of the \texttt{CAReplicaN} reactor.
That reactor sends back the value of the shared variable
at the tag $g_i$, which, in this design, reflects all updates throughout the system
with tags $g_i$ or less.

Note that a read of the value is handled locally on each node.
However, this read will have latency that depends on the time it takes for updates to traverse the network.
In Fig.~\ref{fig:consistencyfirstn}, notice that the reaction to \texttt{query} on line \ref{ln:queryN} is defined \emph{after} the reaction to local and remote updates.
In \lfshort semantics, this ensures that the reaction to \texttt{query} is not invoked at tag $g_i$ until after all local \emph{and remote} updates with tags
$g_i$ or less have been processed. It is this property that gives this program strong consistency.

The key question becomes,
when can the reaction to the \texttt{query} input on line \ref{ln:queryN} of Fig.~\ref{fig:consistencyfirstn}
be executed?  The physical time between $\mathcal{T}(g_i)$ and the time of that reaction invocation becomes
our measure of unavailability.

This scenario matches exactly the scenario leading to the CAL theorem in Section~\ref{sec:cal}.
Hence, the \textbf{unavailability} is given by (\ref{eq:unavailability}).
This result is intuitive.
We will shortly show that the processing offsets can be zero in this case,
so assume $O_i = O_j = 0$.
If execution times are negligible, then $\mathcal{L}_{ij}$ is just the sum of the network latency and clock synchronization error,
and the {unavailability} at node $i$ due to possible updates at node $j$ is
\begin{equation}\label{eq:unav}
\bar{A}_{ij} = \max(L_{ij} + E_{ij}, 0).
\end{equation}
Recall that even though clock synchronization error can be negative,
the above maximization ensures that the unavailability is non-negative.
If we further assume that clock synchronization errors are negligible compared to network latency, then
(\ref{eq:unav}) tells us that the unavailability at $i$ due to possible updates at $j$ is equal to the network latency from $j$ to $i$,
a satisfyingly intuitive result.

We can easily modify the program to improve availability at the cost of consistency.
Specifically, if we replace line \ref{ln:bankconnection} of Fig.~\ref{fig:consistencyfirstn} with this:
\begin{lstlisting}[firstnumber=8,style=framed,language=LF,escapechar=|]
    (bank.publish)+ -> bank.updates after 100 msec;
\end{lstlisting}
then the inconsistency is specified to  be $\bar{C} = 100$ milliseconds.
The \textbf{after} keyword specifies a logical time offset between the sender's tag and the receiver's tag.
In other words, it specifies a \textbf{logical delay} between the initiation of an update by a user and the recording of that update in a state variable of each replica.
This is exactly the tag manipulation considered in Section~\ref{sec:cal}, so, from Theorem~\ref{thm:cal}, the unavailability at node $i$ becomes
\[
\bar{A}_{i} = \max(0, \max_{j \in N} (\mathcal{L}_{ij} - \bar{C}_{ij})) ,
\]
where we have again assumed the processing offsets are zero.
Again, if clock synchronization error and execution times are negligible compared to network latencies,
this states that the unavailability is the largest difference between network latency and logical delays.
With the choice of $\bar{C} = 100$ ms, if the network latency is less than 100 msec, then unavailability becomes zero.
The system can respond instantaneously to user requests.

Even with the logical delay,
this design assures eventual consistency because all updates are applied in the same order at all nodes.
Note that even \emph{local} updates are logically delayed, and hence the same design can be applied even
if the merge operation is not associative and commutative, as in the example in Fig.~\ref{fig:replicated}.

The price for improving availability in this ATM example is  that all queries for the value of the shared variable yield a result that is (logically) 100 msec old.
This means that a query for $x$ may not even reflect a recent \emph{local} update.
If the operations are commutative and associative, however, then local updates need not be delayed.
We can use the structure of Fig.~\ref{fig:consistencyfirst} and apply the logical delay only on the connections that broadcast local updates.
We leave it as an exercise for the reader to modify the programs in Figs.~\ref{fig:consistencyfirst} and \ref{fig:consistencyfirstn}
to accomplish this form of bounded inconsistency by inserting \texttt{after} delays.
If we were to change the design to apply local updates immediately in situations where the merge operation is neither
associative nor commutative, such as the program in Fig.~\ref{fig:replicated}, then we would have
to do some additional work to ensure eventual consistency, using the \textbf{sorted replace} described in Section~\ref{sec:cal}.

What happens if the apparent latency exceeds 100 ms?
There are two possibilities.
We can delay handling of events, thereby increasing unavailability,
or we can proceed with processing events as if the inputs are absent,
thereby increasing inconsistency.
In Section~\ref{sec:implementation}, we describe two coordination mechanisms that we have implemented in \lf,
one of which emphasizes consistency and the other of which emphasizes availability.

Another alternative is to remove the bound on inconsistency altogether while still preserving the property that if the network is repaired, we get eventual consistency.
A one-line change in the \lf program of Fig.~\ref{fig:consistencyfirstn} can realize this strategy.
If we change line \ref{ln:bankconnection} to this subtly different version:
\begin{lstlisting}[firstnumber=8,style=framed,language=LF,escapechar=|]
    (bank.publish)+ |\physicalConn|bank.updates;
\end{lstlisting}
then there is no upper bound on the inconsistency $\bar{C}_{ij}$.
The subtle change is to replace the \textbf{logical connection} \texttt{->} with
a \textbf{physical connection} \physicalConn.
In \lf, this is a directive to assign a new tag $g_i$ at the receiving end $i$
based on a local measurement of physical time $T_i$ when the message is received such that $\mathcal{T}(g_i) = T_i$.
The original tag is discarded.
If all connections between federates are physical connections, then the
federation no longer has any need for clock synchronization.
However, the price we pay is that the order in which updates are applied is now
dependent on apparent latencies.
We preserve eventual consistency only if the merge operation is associative and commutative.
Using physical connections is a draconian measure because it also sacrifices determinacy.
This makes it much harder to define regression tests because a correct execution of the program admits many behaviors.
The use of logical delays, together with the coordination mechanisms given in Section~\ref{sec:implementation}, offers more control.

\subsection{Determinism, Idempotence, and Causal Consistency}\label{sec:determinism}

A \lf program that has only logical connections and no physical connections has deterministic semantics,
in the sense that once tags are assigned, there is exactly one correct execution of the program.
The runtime infrastructure is responsible for ensuring that every reactor is presented with
inputs in tag order, that messages are delivered to reactors exactly once, and that reactions
to messages with identical tags are invoked according to the order specified by the code.  
Moreover, because reactors have explicit input and output ports, \lf has a notion of a communication channel,
a connection between two ports. Each port is guaranteed to have at most one message
at any tag.
These properties, taken together, make it much easier to design consistent distributed programs,
and to trade off consistency against availability.
These properties automatically deliver what
Bailis and Ghodsi call CALM, meaning consistency as logical monotonicity~\cite{BailisGhodsi:13:Consistency}.


\lf also ensures causal consistency, preventing the scenario at $r_5$ in Fig.~\ref{fig:observer}.
The runtime infrastructure, described next in Section~\ref{sec:implementation}, uses the topology of interconnection between reactors together with tags
to ensure that no reaction that reads an input or a state variable is invoked until all precedent reactions have been invoked.
Note that this \emph{does not} require that messages be globally ordered!
It only requires that \emph{each component} (each reactor in \lfshort) see messages in tag order,
and that simultaneous messages (those with the same tag) are handled in precedence order.
We discuss in Section~\ref{sec:implementation} how this is achieved in \lf.

%
%

\section{Implementation} \label{sec:implementation}

We now give two distributed coordination mechanisms, 
which we have implemented as an extension of the \lf coordination language, 
that support arbitrary tradeoffs between consistency and availability as network latency varies.  
With \textbf{centralized coordination}, inconsistency remains bounded by a chosen numerical value 
at the cost that unavailability becomes unbounded under network partitioning.  
With \textbf{decentralized coordination}, unavailability remains bounded by a chosen numerical quantity 
at the cost that inconsistency becomes unbounded under network partitioning. 
Our centralized coordination mechanism is an extension of techniques that have historically been used for distributed simulation, 
an application where consistency is paramount.  
Our decentralized coordination mechanism is an extension of techniques that have been used in distributed databases when availability is paramount.

\subsection{Centralized Coordination} \label{sec:centralized}

Centralized coordination is based on the High-Level Architecture (HLA)~\cite{KuhlEtAl:99:HLA} and other distributed simulation frameworks~\cite{Fujimoto:00:DistributedSimulation,Zeigler:00:Simulation}, with significant extensions that we describe here.
Distributed simulation is a relevant problem because, usually, consistency trumps availability.
A distributed implementation of a simulation is expected to yield the same results as a non-distributed version, only faster.
The HLA is designed for distributed simulation of discrete-event systems, where events have timestamps, and hence addresses a similar problem.

We face two complications, however, that are not present in distributed simulation applications.
The first is that, in our context, unlike simulation, events may materialize out of nowhere with tags derived from the local physical clock.
Our context, in other words, has users \emph{interacting} with the system, and hence availability becomes a concern.
Simulation has no such users.
In \lf, we use physical actions to realize asynchronous stimulus from users.
A second problem is that the programs in Figs.~\ref{fig:consistencyfirstn} and \ref{fig:replicated} have cycles without logical delays,
which are not allowed in HLA.

The HLA, like other distributed simulation frameworks, uses a centralized controller called the \textbf{runtime infrastructure} (\textbf{RTI}).
Each node that wishes to process a tagged event consults with the RTI, which grants permission to advance its \textbf{current tag} to that tag
only when the RTI can assure the node that no event with a lesser tag will later appear.
The existence of physical actions and zero-delay cycles in \lf complicates this assurance and requires extending the protocols used in HLA.
Once a node has advanced its current tag to $g$, it is no longer able to handle any events with tag less than $g$.

Our RTI, like those in distributed simulation frameworks,
realizes a mechanism similar to vector clocks~\cite{LiskovLadin:86:VectorClocks}.
Schwartz and Mattern show that any mechanism that preserves causal consistency
fundamentally has a complexity of at least that of vector clocks~\cite{SchwarzMattern:94:CausalConsistency}.
However, because \lf exposes information about which federates communicate with which,
we have realized significant optimizations.
Our RTI keeps track of the tag to which each federate has advanced, and uses that information, together with network topology information,
to regulate the advancement of the current tag at \emph{downstream} federates based on the activity of their upstream federates.
A federate that has no network inputs, for example, can advance its current tag without consulting the RTI because there is no risk of later seeing
an incoming message that has a tag less than the tag to which it has advanced.
A federate with network inputs, however, must receive an assurance from the RTI, called a \textbf{tag advance grant} (\textbf{TAG}) before it can advance its current tag.

Our first extension over HLA supports zero-delay cycles by introducing a \textbf{provisional tag advance grant} (\textbf{PTAG}), 
where the RTI assures a federate that there will be
no future message with tags \emph{less than} some $g$, but makes no promises about messages with tags \emph{equal to $g$}.
This permits a federate to advance its current tag to $g$ and execute any reactions with no dependence, direct or indirect, on network inputs.
When it has executed such reactions and the next reaction in the reaction sequence depends on network inputs, then the federate is required to block
until it either receives messages on those network inputs or receives an assurance that no message is forthcoming with tag $g$.
Such an assurance is similar to the \textbf{null messages}
of Chandy and Misra~\cite{ChandyMisra:79:DDE}.

In some cases, providing such an assurance is easy. If the upstream federates have all advanced their own current tag beyond $g$,
and have informed the RTI of this fact, then the RTI can provide the required assurance to the downstream federate.
As long as that assurance message is sent along the same order-preserving message channel as tagged messages, then
when a federate receives the assurance, it knows it has received all relevant tagged messages and hence can proceed.
However, if there are cycles between federates that lack logical delays, a federate may need to send a \textbf{null message},
an indicator that no message with tag $g$ is forthcoming,
even before it has completed processing of all events with tag $g$.
It can send such a null message as soon as it has executed or chosen not to execute all reactions that are capable of producing the relevant network output.
Such null messages are similar to those
of Chandy and Misra~\cite{ChandyMisra:79:DDE},
but are only needed in particular circumstances.

When there are physical actions, however, things are still a bit more complicated.
Consider the program in Fig.~\ref{fig:replicated}, focusing particularly on the graphical rendition at the bottom.
This example instantiates four federates, each an instance of the \texttt{Node} reactor class.
Each federate has four input channels on its \texttt{updates} input port.
Under centralized coordination, a federate cannot advance its current tag to $g$ until it receives either a TAG or a PTAG from the RTI with value $g$.
It also cannot advance to $g$ until its physical clock exceeds $\mathcal{T}(g)$ because it has a physical action.

When can the RTI provide a TAG or PTAG message?
This depends on how each federate produces network outputs.
Each federate has a physical action in its \texttt{UserInput} reactor that triggers a network output on the \texttt{publish} port.
The tag $g$ of that output will have $\mathcal{T}(g)$ taken from the local physical clock.
Hence, as soon as the physical clock of a federate exceeds $\mathcal{T}(g)$, downstream federates can be assured that there will be no forthcoming message with timestamp $g$ or less.

Unlike decentralized coordination (see Section~\ref{sec:decentralized} below), centralized coordination does not rely on clock synchronization
except to give a meaning to tags originated by distributed physical actions.
Instead of relying on clock synchronization, in centralized coordination, each federate that has a physical action that can
result in network outputs must notify downstream federates as its physical clock advances.
This is done by periodically sending to the RTI a \textbf{time advance notice} (\textbf{TAN}) message with a time $t$;
this message is a promise to not produce future messages with tags $g$ where $\mathcal{T}(g) \le t$, and hence is also similar to the {null messages}
of Chandy and Misra~\cite{ChandyMisra:79:DDE}.

Unlike Chandy and Misra's technique, in \lf, null messages are only required when communication between federates forms a cycle without logical delays
and when physical actions trigger network outputs.
Unfortunately, all of the applications considered in Section~\ref{sec:example} have such cycles and physical actions and therefore require null messages.
\lf provides a mechanism to control the frequency of the TAN messages, thus controlling the overhead,
but as the frequency of messages decreases, the cost in unavailability increases.
This overhead is avoided in decentralized coordination, explained below in Section~\ref{sec:decentralized},
but at the cost that consistency is sacrificed under network partitioning.

\subsection{Decentralized Coordination} \label{sec:decentralized}

Decentralized coordination extends a mechanism first described by Lamport~\cite{Lamport:84:TimeStamps},
first applied to explicitly timestamped distributed systems in PTIDES~\cite{Zhao:07:PTIDES},
and reinvented at Google to form the core of Google Spanner~\cite{CorbettEtAl:12:Spanner}.
All three of these use timestamps to define the logical ordering of events and physical clocks to determine when it is safe to process
timestamped events.
The physical clocks are assumed to be synchronized with a bound on the clock synchronization error.
All three also assume a bound on network latency.
If these assumptions are met at run time, then all messages will be processed in timestamp order
without any centralized coordination.

Relying on physical clocks has a key advantage with respect to availability.
Specifically, we can assume that even in the presence of complete network partitioning,
physical clocks continue to advance.
If progress is governed by the advancement of physical clocks, then unavailability can be bounded
even with no network connectivity.
This contrasts with centralized coordination, where inconsistency can be bounded, but loss of network connectivity leads to loss
of availability.

\subsubsection{Safe to Advance (STA) Offset.}

In our implementation of decentralized coordination in \lf,
each federate can have an optional \textbf{safe-to-advance} (\textbf{STA}) offset given by the programmer.
The meaning of the STA offset is that if a federate has an earliest pending event with tag $g$,
then it can advance its current tag to $g$ when current physical time $T$ satisfies $T \ge \mathcal{T}(g)+ \text{STA}$.
Hence, to handle a user request that gets assigned tag $g$, the federate needs to wait at least
until physical time exceeds $\mathcal{T}(g)$ by the STA offset.
Put another way, the STA offset is a time interval beyond $\mathcal{T}(g)$ that a federate needs to wait before it
can assume that it will not later receive any input messages with tags less than $g$.
By default, STA = 0.
The STA offset is closely related to the safe-to-process offset of Zhao et al.~\cite{Zhao:07:PTIDES},
but is more provisional.
It gives a time threshold for \emph{committing} to a tag advance, but not necessarily fully \emph{processing} that tag advance.
Put another way, it gives a time threshold at which the federate can assume it has seen all messages with tags
\emph{less than} $g$, but it cannot necessarily assume it has seen all messages with tags
\emph{equal to} $g$.
This distinction turns out to be important for the replicated data store examples we have seen.

Obviously, the STA offset affects availability and is clearly closely related to the processing offset of Definition~\ref{def:processing}.
There is, however, a subtle but important distinction.
The processing offset of Definition~\ref{def:processing} is a property of a \emph{trace}, an actual execution,
whereas the STA is a \emph{specification}.
The \lf code generator generates code where, when executed, every trace will have the property that
\begin{equation}\label{eq:offset}
O_i \ge \text{STA}_i
\end{equation}
for each federate $i$.
If the STA offset is not sufficiently large for a particular program,
then the consistency requirements of the program will not be met.
Our task, therefore, is to determine sufficiently large STA offsets such that,
if the observed apparent latencies are within our assumed constraints,
the program will process all events in tag order, thereby achieving the desired consistency.
Only when the apparent latencies exceed our assumed constraints will the program sacrifice consistency in order to maintain availability.

The STA offsets depend on assumed bounds on apparent latency,
and vice versa, the assumed bounds on apparent latency depend on the STA offsets,
which brings us to a second subtlety.
In Definition~\ref{def:apparent},
apparent latency is also a property of a {trace}, whereas, to use it to derive
the STA offsets, we need to use it as a bound on \emph{all reasonable traces}.
Any assumed bound may be exceeded in practice (e.g., the network becomes partitioned),
and the strategy of decentralized coordination is to sacrifice consistency rather than availability when this occurs.
This condition will be detectable, and \lf supports specification of \textbf{fault handlers} for such conditions (see
Section~\ref{sec:fault}).

A third subtlety is that, in Definition~\ref{def:apparent}, the apparent latency is a property of a pair of \emph{processes},
sequential procedures where one sends updates to another.
\lf, however, is a more richly structured language.
Federates themselves may be concurrent, running in parallel on multicore machines, for example,
and communication between federates is mediated by input and output \textbf{ports} that, pairwise,
give specific communication channels over which messages with monotonically increasing tags flow.
As a consequence, apparent latency between one federate and another may vary depending on which
communication channel between the two is used.
Moreover, each pair of send-receive ports may have a different logical delay.

\subsubsection{Causality.}

To analyze a \lf program, we need to redo that analysis of Section~\ref{sec:cal} using the structure of the program.
Specifically, it is possible to tell by looking at the program whether an event at one port can result in an event at another port,
and we can find bounds on the relationship between the tags of these two events.
This analysis must be done carefully, however, because we have to distinguish whether an event at one port
can \emph{cause} an event at another from whether it can \emph{influence} an event at another port.

To determine whether a message with a certain tag can exist, we need to analyze the
\textbf{counterfactual causality} properties of the \lf program.
Counterfactual causality~\cite{pearl2009causality} is a relation between events $e_1$ and $e_2$ where
$e_2$ would \emph{not occur} were it not for the occurrence of $e_1$.
We distinguish this from \textbf{causal influence}, where event $e_1$ can \textbf{causally affect} $e_2$~\cite{Lamport:78:Time}.
In a \lf program, any input to a reactor with tag $g$ can causally affect any output with tag larger than $g$ and some outputs with
tag equal to $g$ because a reaction to that input can change the state of the reactor.
However, only some inputs counterfactually cause particular outputs, which then in turn
counterfactually cause other inputs.

Specifically, consider a reactor like this:
\begin{lstlisting}[style=framed,language=LF,escapechar=|]
reactor DirectFlowThrough {
    input in:int;
    output out:int;
    reaction(in) -> out {=
        ...
    =}
}
\end{lstlisting}
Because of the reaction signature, we assume that an event at input named \texttt{in} can counterfactually cause an event
at the output named \texttt{out}.
We do not need to analyze the body of the reaction (which is written in the target language) to determine this fact.
In contrast, consider:
\begin{lstlisting}[style=framed,language=LF,escapechar=|]
reactor CausalInfluence {
    input in1:int;
    input in2:int;
    output out:int;
    state s:int(0);
    reaction(in1) {=
        ...
    =}
    reaction(in2) -> out {=
        ...
    =}
}
\end{lstlisting}
Here, input \texttt{in1} causally influences output \texttt{out} (because the first reaction can change the state, and
the second reaction can use that updated state), but it does not counterfactually cause the output.
For the output to occur, a message must arrive on \texttt{in2}.

In the above example, an input with tag $g$ on \texttt{in1} can causally influence any output with tag $g$ or larger.
If the reactions were given in the opposite order, then it would only be able to causally influence an output with tag \emph{larger than} $g$.
This is because, given simultaneous inputs, reactions of a reactor are invoked in the order that they are declared.
This distinction proves important when analyzing the distributed replicated databases considered earlier.

Using actions, a reactor can declare a logical delay:
\begin{lstlisting}[style=framed,language=LF,escapechar=|]
reactor IndirectFlowThrough {
    input in:int;
    output out:int;
    logical action a:int(10 msec);  // minimum delay of 10 msec.
    reaction(a) -> out {=
        ...
    =}
    reaction(in) -> a {=
        ...
    =}
}
\end{lstlisting}
In this example, the \textbf{logical action} has a \textbf{minimum delay} property (set to 10 msec).
The pair of reactions, taken together, reveal that the input with tag $g$ can counterfactually cause an output
with tag $g'$, where $\mathcal{T}(g')$ is larger than $\mathcal{T}(g)$ by at least 10 msec.
This introduces a logical delay on the path from \texttt{in} to \texttt{out}.
\clearpage
Consider:
\begin{lstlisting}[style=framed,language=LF,escapechar=|]
reactor Composition {
    a = new IndirectFlowThrough();
    b = new DirectFlowThrough();
    a.out -> b.in;
}
\end{lstlisting}
Because of the connection, we can infer that output \texttt{a.out} can counterfactually cause input \texttt{b.in}
with no logical delay.
Moreover, because of the minimum delay property,
we can infer that input \texttt{a.in} can counterfactually cause input \texttt{b.in}
with logical delay of at least 10 msec.

The connection may also have a logical delay (written with the \textbf{after} keyword), as in:
\begin{lstlisting}[style=framed,language=LF,escapechar=|,firstnumber=4]
    a.out -> b.in after 20 msec;
\end{lstlisting}
Now, the program reveals that input \texttt{a.in} can counterfactually cause input \texttt{b.in}
with logical delay of at least 30 msec. 

\subsubsection{Safe-to-Assume Absent (STAA).}

Similar to the STA offset (which is, essentially, found in Lamport, PTIDES, and Spanner), we extended \lf to allow specification
of a \textbf{safe-to-assume-absent} (\textbf{STAA}) offset associated with a network input port.
The STAA offset is used to constrain when a reaction that depends on an input port can be invoked.
Specifically, it asserts that the invocation of any reaction at tag $g$ that depends, directly or indirectly, on a network input port $p_i$
is delayed until either an input is received on port ${p_i}$ with tag $g$
or the physical clock $T_i$ at $i$ satisfies
\begin{equation}\label{eq:staadef}
T_i \ge \mathcal{T}(g) + \text{STA}_i + \text{STAA}_{p_i}.
\end{equation}
At this physical time, federate $i$ assumes it has seen all inputs at port $p_i$ with tags less than \emph{or equal} to $g$
(vs. the STA offset alone, when it can assume it has seen all inputs with tags less than $g$).
If no message has arrived with tag $g$, the federate assumes there is no message with tag $g$.
It would be an error, to be handled as a fault condition, to later receive a message with tag $g$.

A positive STAA offset causes the federate to block execution of reactions in the relevant reactor until either
physical time advances sufficiently or a message arrives.
In this circumstance, a null message could be used to reduce the amount of blocking,
but, unlike with centralized coordination, no null message is required to make progress. It is sufficient for physical time to advance.
Using a null message would just be an optimization that may allow progress sooner.


As we will see below, the STA and STAA offsets together ensure that any event that causally influences another
is processed first. To determine the STA, we need to consider causal influence, but to determine STAA,
we only need to consider counterfactual causality.

\subsubsection{Decentralized Coordination for the Replicated Data Store.}

Consider now the replicated data store in Fig.~\ref{fig:replicated}.
The pessimistic analysis of Section~\ref{sec:pessimistic}, for this program,
yields an infinite processing offset for all federates.
We will now show that, by leveraging the semantics of \lf and our extensions to its runtime,
the program can be executed correctly with finite STA and STAA offsets.
We show how to determine these offsets for each of the federates and their input ports.

\begin{figure}[tb]
\begin{lstlisting}[style=framed,language=LF,escapechar=|]
target C {
    coordination: decentralized
}

import UserInput from "UserInput2.lf";
import ReplicaN from "ReplicatedDataStore.lf";

federated reactor (
    N:int(4)
) {
    u = new[N] UserInput();
    r = new[N] ReplicaN(N = N);
    u.update -> r.query;
    r.response -> u.current_value;
    (u.update)+ -> r.updates;
}
\end{lstlisting}
\includegraphics[width=\linewidth]{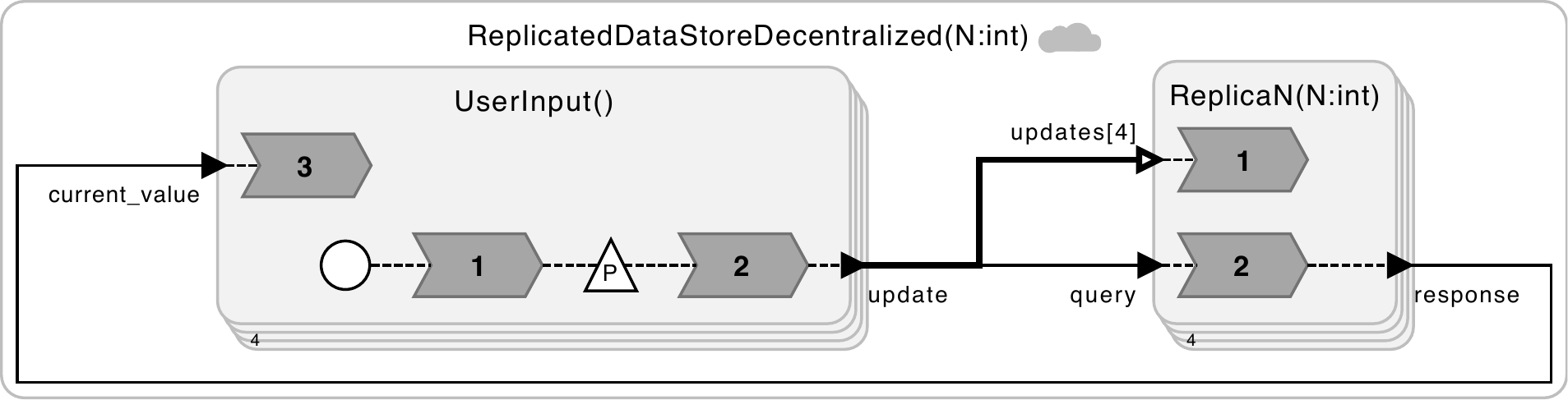}
\caption{Version of the replicated data store of Fig.~\ref{fig:replicated} suitable for decentralized coordination, where \texttt{UserInput} and \texttt{ReplicaN} have been split into separate federates so that they can independently advance their current tags. \label{fig:replicateddecentralized}}
\end{figure}

First, it will shortly become obvious that we need to separate \texttt{ReplicaN} and \texttt{UserInput} into distinct federates,
even if they run on the same host, so that they can independently advance their current tags.\footnote{In the current implementation
of \lf, an entire federate, with all its component reactors, advances the current tag together. In principle,
some future implementation of \lf could allow component reactors to independently advance their current tags.
This could be accomplished using mechanisms similar to \lfshort's federated execution.
But for now, the only available mechanism to permit independent advancement of tags is to separate the reactors into distinct federates.
Sometimes, however, it is not possible to create such a separation.
The \lf code generator assumes that any two reactions of the same reactor share state.
It does not analyze the target code to check whether this is the case.
As a consequence, if the reactions of \texttt{UserInput} and \texttt{ReplicaN} in Fig.~\ref{fig:replicated} were
instead reactions of the same reactor, then it would not be possible to separate them
into distinct federates nor to independently advance their tags.
}
The refactored program is shown in Fig.~\ref{fig:replicateddecentralized}.
Given such a separation, note that each of the four instances of \texttt{ReplicaN} receive inputs from each of the four instances of
\texttt{UserInput}, including the one running on the same host.

\begin{figure}[tb]
\centering
\includegraphics[width=0.6\linewidth]{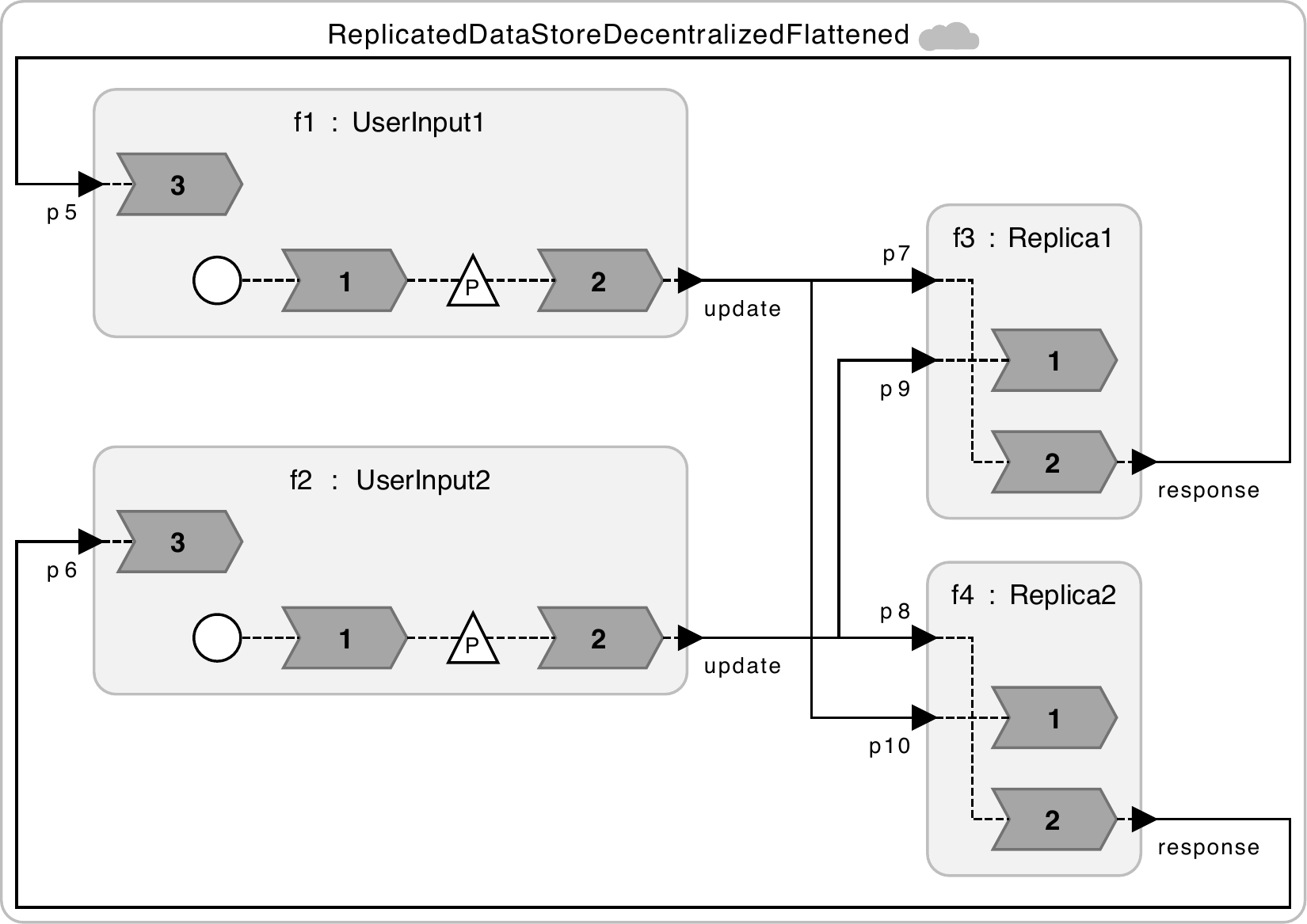}
\caption{Minimal version of the replicated data store of Fig.~\ref{fig:replicateddecentralized} with ports
renamed to all be unique. \label{fig:replicateddecentralizedflattened}}
\end{figure}

To further simplify the explanation, we have reduced the program to that shown in Fig.~\ref{fig:replicateddecentralizedflattened},
which is a minimal version that avoids the compact bank and multiport syntax of \lf and renames all the ports so that
they have unique names.
There are a total of four federates and six input ports now, so we need to determine four STA offsets and six STAA offsets.

A key property of our execution policy for \lf programs is a federate advances its current tag to $g$ only after it has
completed handling all events with lesser tags,
and then it completes handling of all events with tag $g$ before advancing to another larger tag.
I.e., even if the federate is executing reactions concurrently (e.g., on a multicore machine), it
performs a barrier synchronization with each tag advance.
There is no such barrier synchronization across federates, but we need the barrier synchronization within
a federate, as will become obvious.

Consider first federate $f_1$, the \texttt{UserInput} at the upper left of Fig.~\ref{fig:replicateddecentralizedflattened}.
Suppose that the physical action (depicted as a triangle with a ``P'') triggers and is assigned tag $g_1$ using the local physical clock.
The question now is, when can the federate advance its tag to $g_1$?
It has to ensure that it has seen all inputs with lesser tags, including events that may have been sent to port $p_5$.
For the first triggering of the physical action, it is evident from the program structure that there is no event at $p_5$ with
a lesser tag because all events at $p_5$ are ultimately counterfactually caused by this same physical action.
Therefore, the federate can safely advance to tag $g_1$ and invoke its reaction 2 with no delay.
Thus, it seems that $STA_1 = 0$ could work for federate $f_1$, at least for this first event.

Once federate $f_1$ has advanced to $g_1$, it will block any further advances until physical time
advances past $\mathcal{T}(g_1) + \text{STA}_1 + \text{STAA}_{p_5}$.
Assuming it does this correctly, then for the next triggering of the physical action with tag $g_2 > g_1$,
the federate will not even face this question of whether to advance its tag to $g_2$
until it has completed processing events with tag $g_1$.
More generally, each time the physical action triggers with tag $g_n$, $n > 1$,
by the time the federate is considering advancing its tag to $g_n$, it
will have completed processing all events with tag $g_{n-1}$.
Therefore, it does not need to wait for physical time to further advance.
Hence, by induction, $\text{STA}_1 = 0$ for all tag advances.
The same argument applies for federate $f_2$, yielding
\begin{eqnarray}
\text{STA}_1 &=& 0\\
\text{STA}_2 &=& 0 .
\end{eqnarray}

To determine $\text{STAA}_{p_5}$, we can follow all physical time lags that might occur
on the path from the original source.
We now make a critical assumption that is required to ensure finite offsets:
\begin{assumption}\label{ass:critical}
Reaction 2 of $f_1$ is invoked exactly at $\mathcal{T}(g_1) + \text{STA}_1$ or negligibly thereafter.
\end{assumption}
We will see in Section~\ref{sec:exavailability} how this assumption can be enforced by the \lf program, but first, we determine the consequences
of this assumption. With it, any $\text{STAA}_{p_5}$ that satisfies the following will suffice:
\[
\text{STAA}_{p_5} \ge \text{STA}_3 + \max(\text{STAA}_{p_7}, \text{STAA}_{p_9}) + X_{31} + X_{32} + L_{13} + E_{13},
\]
where $X_{ij}$ is an execution time bound on reaction $j$ of federate $f_i$, $L_{ij}$ is a communication latency bound on
messages from $f_j$ to $f_i$, and $E_{ij}$ is a bound on the clock synchronization error from $f_j$ to $f_i$.
The maximization and the presence of $X_{31}$ is a consequence of the \lf semantics that requires that if reactions 1 and 2
of the same reactor
are both enabled at any tag $g$, then reaction 1 must run to completion before reaction 2 is invoked.

Let us make a simplifying assumption to manage the complexity of this
(this assumption, unlike Assumption~\ref{ass:critical}, is not necessary, but drastically simplifies our example).
Specifically, let's assume that execution time bounds are negligible compared to communication latencies.
With this assumption, we get
\begin{equation}\label{eq:p5}
\text{STAA}_{p_5} \ge \text{STA}_3 + \max(\text{STAA}_{p_7}, \text{STAA}_{p_9}) + L_{13} + E_{13}.
\end{equation}
We can write a similar inequality for $\text{STAA}_{p_6}$,
\begin{equation}\label{eq:p6}
\text{STAA}_{p_6} \ge  \text{STA}_4 + \max(\text{STAA}_{p_8}, \text{STAA}_{p_{10}}) + L_{24} + E_{24}.
\end{equation}
So far, we have $\text{STA}_1 = \text{STA}_2 = 0$ and these two inequalities.
Let us now look at $\text{STA}_3$.

The question is, given an event with tag $g_3$ that federate $f_3$ wishes to process,
how much physical time should it wait before advancing to tag $g_3$?
First, this federate has no local sources of events (actions or timers), so the event must be an
input on either port $p_7$ or $p_9$.  In either case, in order to advance to $g_3$, the federate needs
to be assured that it has seen all inputs earlier than $g_3$ \emph{on the other port} in order to ensure causality.
(\lf assumes that messages on each channel are delivered in tag order.)

If the input has arrived on $p_7$, then it requires
\[
 \text{STA}_3 \ge  \text{STA}_2 + X_{22} + L_{32} + E_{32},
\]
which is obtained by following the counterfactual causality chain upstream from $p_9$.
Using $ \text{STA}_2 = 0$ and the negligible execution time assumption,
\[
 \text{STA}_3 \ge  L_{32} + E_{32}.
\]

If the input with tag $g_3$ has arrived instead on $p_9$, then we require
\begin{equation}\label{eq:sta3}
 \text{STA}_3 \ge  \text{STA}_1 + X_{12} + L_{31} + E_{31}.
\end{equation}
Combining these and ignoring execution times, we get
\[
 \text{STA}_3 \ge \max( L_{32} + E_{32},  L_{31} + E_{31}).
\]
There are no further constraints on $\text{STA}_3$, so we can simply set
\begin{equation}
 \text{STA}_3 = \max( L_{32} + E_{32},  L_{31} + E_{31}).
\end{equation}
Similarly,
\begin{equation}
 \text{STA}_4 = \max( L_{41} + E_{41},  L_{42} + E_{42}).
\end{equation}

Similar reasoning leads to
\begin{eqnarray*}
\text{STAA}_{p_7} &=& L_{32} + E_{32} -  \text{STA}_3 = \min(0,  L_{32} + E_{32} -  L_{31} - E_{31}) \\
\text{STAA}_{p_9} &=& L_{31} + E_{31} -  \text{STA}_3 = \min(0,  L_{31} + E_{31} -  L_{32} - E_{32}) \\
\text{STAA}_{p_8} &=& L_{41} + E_{41} -  \text{STA}_4 = \min(0,  L_{41} + E_{41} -  L_{42} - E_{42}) \\
\text{STAA}_{p_{10}} &=& L_{42} + E_{42} -  \text{STA}_4 = \min(0,  L_{42} + E_{42} -  L_{41} - E_{41}).
\end{eqnarray*}
There is no point in having a negative STAA offset, so
\begin{equation}
\text{STAA}_{p_7} =
\text{STAA}_{p_9} =
\text{STAA}_{p_8} =
\text{STAA}_{p_{10}} = 0.
\end{equation}
Finally, from (\ref{eq:p5}) and  (\ref{eq:p6}), we can set
\begin{eqnarray}
\text{STAA}_{p_5} &=&  \text{STA}_3 + \max(\text{STAA}_{p_7}, \text{STAA}_{p_9}) + L_{13} + E_{13}. \nonumber \\
&=& \max(L_{32} + E_{32}, L_{31} + E_{31}) + L_{13} + E_{13}\\
\text{STAA}_{p_6} &=& \text{STA}_4 + \max(\text{STAA}_{p_8}, \text{STAA}_{p_{10}}) + L_{24} + E_{24} \nonumber \\
&=& \max(L_{41} + E_{41}, L_{42} + E_{42}) + L_{24} + E_{24}.
\end{eqnarray}

As a sanity check, let's simplify further by assuming that $f_1$ and $f_3$ are mapped to the same host,
so that $E_{31} = E_{13} = E_{42} = E_{24} = 0$, and $L_{31}$, $L_{13}$, $L_{42}$ and $L_{24}$ are all negligible.
Under these assumptions, we get the following total results:
\begin{eqnarray*}
\text{STA}_1 &=& 0 \\
\text{STA}_2 &=& 0 \\
\text{STA}_3 &=& \max(L_{32} + E_{32}, 0) \\
\text{STA}_4 &=& \max(L_{41} + E_{41}, 0) \\
\text{STAA}_{p_5} &=&  \max(L_{32} + E_{32}, 0) \\
\text{STAA}_{p_6} &=& \max(L_{41} + E_{41}, 0) \\
\text{STAA}_{p_7} &=& 0 \\
\text{STAA}_{p_9} &=& 0 \\
\text{STAA}_{p_8} &=& 0 \\
\text{STAA}_{p_{10}} &=& 0. \\
\end{eqnarray*}
Further, let's assume that clock synchronization error is negligible compared to network latencies.
Then, we get:
\begin{eqnarray}\label{eq:simple}
\text{STA}_1 &=& 0 \nonumber \\
\text{STA}_2 &=& 0 \nonumber \\
\text{STA}_3 &=& L_{32} \nonumber \\
\text{STA}_4 &=& L_{41} \nonumber \\
\text{STAA}_{p_5} &=& L_{32} \nonumber \\
\text{STAA}_{p_6} &=& L_{41} \nonumber \\
\text{STAA}_{p_7} &=& 0 \nonumber \\
\text{STAA}_{p_9} &=& 0 \nonumber \\
\text{STAA}_{p_8} &=& 0 \nonumber \\
\text{STAA}_{p_{10}} &=& 0.
\end{eqnarray}
These results are intuitive.
They show that, at the \texttt{UserInput} federates, when a physical action triggers with tag $g$,
the federate can immediately advance its current tag to $g$, so reaction~2 can be immediately invoked,
resulting in a network output.
Whether to invoke reaction~3 the \texttt{UserInput} federates
cannot be determined until physical time exceeds $\mathcal{T}(g)$ by a bound on the
network latency from the other host,
a satisfyingly intuitive result because that is where a remote update may occur.

At the \texttt{Replica} federates, when they receive an input with tag $g$, they can advance
to tag $g$ only when physical time exceeds $\mathcal{T}(g)$ by the bound on the network latency from
the other host.  This too is intuitive because only at that physical time can they be sure there
is no forthcoming message from the other host with a lesser tag.

For this particular example, 
at the \texttt{UserInput} federates,
as long as the assumptions on network latency and clock synchronization are satisfied,
there \emph{will} be a network input with tag $g$, and reaction~3 will be invoked.
But the \lfshort infrastructure cannot be sure that this is the case without imposing further constraints on the target
code in reaction~2 of \texttt{UserInput} and reaction~2 of \texttt{Replica}.
Those reactions are free to choose to not produce an output.

As of this writing, in \lf, the STA and STAA offsets must be derived by hand and provided as part of the specification of the program.
We leave it to future work to derive these thresholds automatically given assumptions about apparent latency.
This will require performing analysis of the structure of the program and rejecting programs that result in infinite values
for these offsets.
The analysis is simple for this program, but it could be quite challenging in general.
For example, if \texttt{UserInput} had a second physical action
and there were a logical delay $D$ somewhere along the path from its \texttt{update} output back to its \texttt{current\_value} input,
then $\text{STA}_i = 0$ is not necessarily any longer valid.

Why did we have to separate \texttt{UserInput} and \texttt{Replica}
into distinct federates?
Were they in the same federate, then 
$ \text{STA}_3 =  \text{STA}_1$.
From (\ref{eq:sta3}), we have the constraint that $\text{STA}_3 > \text{STA}_2$.
Correspondingly, $\text{STA}_4 > \text{STA}_1$ and $\text{STA}_4 = \text{STA}_2$.
Combining these, we get $\text{STA}_1 > \text{STA}_2 > \text{STA}_1$,
a constraint which is not satisfiable.

\subsection{Unavailability in the Replicated Data Store}\label{sec:exavailability}

Using the most simplified result, given by (\ref{eq:simple}), we can see the consequences of the CAL theorem
for the replicated data store example.
This program is strongly consistent.
There are no logical delays, so each replica will agree on the value of the shared variable at every tag.
A user who issues a query gets a reply when reaction~3 of \texttt{UserInput} is invoked.
From (\ref{eq:simple}), we see that $\text{STAA}_{p_5} = L_{32}$ and
$\text{STAA}_{p_6} = L_{41}$, which means that the time it takes to respond to a user query
is at most the network latency between nodes.

This is intuitive and not surprising.
However, there is more subtle consequence.
Recall Assumption~\ref{ass:critical}, that
reaction 2 of $f_1$ is invoked exactly at $\mathcal{T}(g_1) + \text{STA}_1$ or negligibly thereafter.
Because of the barrier synchronization for advancement of the current tag in a federate,
this assumption \emph{may not be met} if the physical action triggers too closely after
its previous trigger, specifically within $L_{32}$.
If the physical action triggers while \texttt{UserInput1} is waiting on port $p_5$, then there
will be a delay in the invocation of reaction 2 and the derived STA and STAA offsets are no longer
assured to be valid. A fault condition may occur.

Fortunately, \lf provides mechanisms to prevent such eventualities.
First, a physical action can have a \textbf{minimum spacing} parameter,
a minimum logical time interval between tags assigned to events.
When the environment tries to violate this constraint by issuing requests too quickly,
the programmer can specify one of three policies: drop, replace, or defer.
The drop policy simply ignores the event.
The replace policy replaces any previously unhandled event, or if the event has already been handled, defers.
The defer policy assigns at tag $g$ to the event with timestamp $\mathcal{T}(g)$ that is larger than the previous event by the specified minimum spacing.
This feature of the language can be used to help protect a system against denial of service attacks
that might otherwise trigger fault conditions.

While the minimum spacing parameter ensures that tags are sufficiently spaced, it does not,
by itself, ensure that the scheduler will prioritize execution of reaction~2 so as to satisfy Assumption~\ref{ass:critical}.
\lf provides a mechanism to ensure this, a \textbf{deadline} that can be associated with a reaction.
The syntax for this is as follows:
\begin{lstlisting}[firstnumber=38,style=framed,language=LF,escapechar=|]
   physical action r;
   reaction(r) {=
      ... normal case
   =} deadline(1 msec) {=
      ... exception case
   =}
\end{lstlisting}
The semantics of a \lf \texttt{deadline} is that if the reaction to an event with tag $g$ is invoked at a physical time
$T > \mathcal{T}(g)$, then instead of invoking the ``normal case'' reaction,
the ``exception case'' reaction will be invoked.
This provides a mechanism to handle overload conditions, but, more importantly, the
deadline provides a hint to the scheduler to prioritize invocation of this reaction.
Indeed, \lf uses an earliest-deadline-first (EDF) scheduling policy, thereby ensuring, for sufficiently
simple programs, that Assumption~\ref{ass:critical} will be met.

\subsection{Handling Fault Conditions}\label{sec:fault}

Whether we use centralized or decentralized coordination, faults can occur.
Centralized coordination bounds inconsistency, and when faults make it impossible to proceed without exceeding those bounds,
program execution must pause or stop.
In this case, the RTI is a single point of failure, so a fault-tolerant system will need to have a mechanism to elect a new RTI when
one fails.
Progress at a federate can also be stopped by failure of an upstream federate.
A mechanism for restarting federates could help mitigate this risk, but the time it takes to perform such a restart will
inevitably decrease availability.
A network partition will also stop progress.
If communication from a federate to the RTI is lost, no progress is possible either for the federate or
for federates downstream of it.
There is no mitigation for this loss of availability that keeps inconsistency bounded, as shown by the CAL theorem.

When we want to bound unavailability instead of inconsistency, we should use decentralized coordination.
Decentralized coordination may also be more efficient, because it does not require null messages to handle physical actions,
but it requires clock synchronization, which also increases network traffic.

For decentralized coordination to work, we need to assume a bound on apparent latency $\mathcal{L}$, which means bounding
network latency, execution time overhead, and clock synchronization error.
\lf provides a convenient mechanism for handling faults where these assumptions have been violated.
Specifically, in \lfshort, the program can associate with each reaction that is triggered by a network input
a fault handling reaction that will be invoked instead of the regular reaction whenever a network input arrives that has
an unexpectedly early tag such that the local current tag has already advanced past that tag.
In this case, there is no way to correctly execute the program.
What the fault handler does is application dependent.
A database application, such as Spanner, may, for example, overlay a transaction schema on top of the mechanisms
provided by decentralized coordination and reject a transaction when such a fault occurs.
A bulletin board application, in contrast, could provide a more graceful recovery that simply re-orders the posts.

Decentralized coordination guarantees a bound on unavailability.
The \lf program specifies a bound on inconsistency, but the CAL theorem tells us that if apparent latency
exceeds some threshold, the bound on inconsistency can no longer be achieved.
If and when this happens, if specified, a fault handler will be invoked.

Centralized coordination, in contrast, guarantees a bound on inconsistency.
Is there a way in this case to specify a bound on unavailability?
In \lf, the programmer can specify a \textbf{deadline} for a particular reaction invocation.
Suppose we replace lines \ref{ln:reactionbalance} through \ref{ln:reactionbalancen} in Fig.~\ref{fig:user} with this:
\begin{lstlisting}[firstnumber=38,style=framed,language=LF,escapechar=|]
   reaction(balance) {=
      printf("Balance: %d\n", balance->value);
   =} deadline(100 msec) {=
      printf("Apologies for the delay! Your balance is %d\n", balance->value);
   =}
\end{lstlisting}
The semantics of a \lf \texttt{deadline} is that if the reaction to an event with tag $g$ is invoked at a physical time $T$
that exceeds the logical time $\mathcal{T}(g)$ by more than 100 milliseconds, then the
second body of code will be invoked instead of the first.
This deadline can be interpreted as a specified bound on unavailability.
However, the \lfshort \texttt{deadline} will detect a violation only when the
\texttt{balance} input finally appears.
In the presence of a network partition, that will not occur until the network is repaired.
While this is better than nothing, we leave it as further work to find an extension to the language that can
detect earlier a loss of availability exceeding a specified threshold.

\section{Conclusions}\label{sec:conclusion}

Our generalization of Brewer's CAP theorem, which we call the CAL theorem,
quantifies the relationship between inconsistency, unavailability, and apparent latency,
where apparent latency includes network latency, execution time overhead, and clock synchronization error.
We have shown how the \lf coordination language enables arbitrary tradeoffs between consistency and availability as apparent latency varies.
We have extended the implementation of \lf with two forms of coordination for distributed programs.
With \emph{centralized coordination}, inconsistency remains bounded by a chosen numerical value 
at the cost that unavailability becomes unbounded under network partitioning.  
With \emph{decentralized coordination}, unavailability remains bounded by a chosen numerical quantity 
at the cost that inconsistency becomes unbounded under network partitioning. 
In both cases, \lf semantics provides predictable and repeatable behaviors in the absence of faults.
In the case of decentralized coordination, a simple fault handling mechanism is quite general and enables
an application to react in controlled ways to loss of consistency while preserving availability.
For centralized coordination, a deadline violation handler serves as a fault handler for loss of availability
while preserving consistency.

Both coordination mechanisms given here are significant extensions over prior art.
Our centralized coordination extends previous methods that have been used for distributed simulation
to support asynchronous injection of user-input events and cycles in the communication topology.
Specifically, it adds to the techniques of HLA a time-advance notice (TAN) and a provisional time-advance grant (PTAG).
Our decentralized coordination extends previous methods used for distributed databases with a safe-to-assume-absent (STAA)
offset that enables better support for cyclic communication structures and asynchronously injected user events.

We have given concrete realizations of programs illustrating the tradeoffs.
These realizations are complete executable \lf programs.
The \lfshort compiler, \texttt{lfc}, or its Eclipse-based integrated development environment, Epoch,
translate these programs into standalone C programs that will run on any POSIX-compliant platform.

\section*{Acknowledgments}
The authors thank I\~nigo Incer for helpful suggestions on an earlier version.

The work in this paper was supported in part by the National Science Foundation (NSF) award \#CNS-1836601 (Reconciling Safety with the Internet) and the iCyPhy (Industrial Cyber-Physical Systems) research center, supported by Denso, Siemens, and Toyota.

%
%
%
\bibliographystyle{splncs04}
\bibliography{Refs}
%
%
%
%
%
\end{document}